\documentstyle[12pt]{article}
\begin{document}

\def\ds{\displaystyle}
\def\beq{\begin{equation}}
\def\eeq{\end{equation}}
\def\bea{\begin{eqnarray}}
\def\eea{\end{eqnarray}}
\def\beeq{\begin{eqnarray}}
\def\eeeq{\end{eqnarray}}
\def\ve{\vert}
\def\vel{\left|}
\def\ver{\right|}
\def\nnb{\nonumber}
\def\ga{\left(}
\def\dr{\right)}
\def\aga{\left\{}
\def\adr{\right\}}
\def\lla{\left<}
\def\rra{\right>}
\def\rar{\rightarrow}
\def\nnb{\nonumber}
\def\la{\langle}
\def\ra{\rangle}
\def\ba{\begin{array}}
\def\ea{\end{array}}
\def\tr{\mbox{Tr}}
\def\ssp{{\Sigma^{*+}}}
\def\sso{{\Sigma^{*0}}}
\def\ssm{{\Sigma^{*-}}}
\def\xis0{{\Xi^{*0}}}
\def\xism{{\Xi^{*-}}}
\def\qs{\la \bar s s \ra}
\def\qu{\la \bar u u \ra}
\def\qd{\la \bar d d \ra}
\def\qq{\la \bar q q \ra}
\def\gGgG{\la g^2 G^2 \ra}
\def\q{\gamma_5 \not\!q}
\def\x{\gamma_5 \not\!x}
\def\g5{\gamma_5}
\def\sb{S_Q^{cf}}
\def\sd{S_d^{be}}
\def\su{S_u^{ad}}
\def\ss{S_s^{??}}
\def\sbp{{S}_Q^{'cf}}
\def\sdp{{S}_d^{'be}}
\def\sup{{S}_u^{'ad}}
\def\ssp{{S}_s^{'??}}
\def\sig{\sigma_{\mu \nu} \gamma_5 p^\mu q^\nu}
\def\fo{f_0(\frac{s_0}{M^2})}
\def\ffi{f_1(\frac{s_0}{M^2})}
\def\fii{f_2(\frac{s_0}{M^2})}
\def\O{{\cal O}}
\def\sl{{\Sigma^0 \Lambda}}
\def\es{\!\!\! &=& \!\!\!}
\def\ap{\!\!\! &\approx& \!\!\!}
\def\ar{&+& \!\!\!}
\def\ek{&-& \!\!\!}
\def\kek{\!\!\!&-& \!\!\!}
\def\cp{&\times& \!\!\!}
\def\se{\!\!\! &\simeq& \!\!\!}
\def\eqv{&\equiv& \!\!\!}
\def\kpm{&\pm& \!\!\!}
\def\kmp{&\mp& \!\!\!}

% .........................................................

\def\simlt{\stackrel{<}{{}_\sim}}
\def\simgt{\stackrel{>}{{}_\sim}}

% .........................................................

\title{
         {\Large
                 {\bf
Double--lepton polarization asymmetries in 
$\Lambda_b \rar \Lambda \ell^+ \ell^-$ decay
                 }
         }
      }

\author{\vspace{1cm}\\
{\small T. M. Aliev \thanks
{e-mail: taliev@metu.edu.tr}\,\,,
V. Bashiry% \thanks
%{e-mail: e114288@metu.edu.tr}
\,\,,
M. Savc{\i} \thanks
{e-mail: savci@metu.edu.tr}} \\
{\small Physics Department, Middle East Technical University,
06531 Ankara, Turkey} }

\date{}

\begin{titlepage}
\maketitle
\thispagestyle{empty}

\begin{abstract}
Double--lepton polarization asymmetries in
$\Lambda_b \rar \Lambda \ell^+ \ell^-$ decay are calculated using a general, 
model independent form of the effective Hamiltonian. The sensitivities of 
these asymmetries to the new Wilson coefficients are studied in detail. 
Furthermore, the correlations between averaged double--lepton polarization 
asymmetry and branching ratio are analyzed. It is shown that there exist 
certain regions of the new Wilson coefficients where new physics can be 
established by measuring the double--lepton polarization asymmetries only.
\end{abstract}

%\vspace{1cm}
~~~PACS numbers: 12.60.--i, 13.30.--a. 13.88.+e
\end{titlepage}

\section{Introduction}

Rare B--decays induced by the flavor--changing neutral current (FCNC) 
$b \rar s(d) \ell^+ \ell^-$ have received a lot of theoretical interest 
\cite{R6601}. These transitions provide an 
important consistency check of the standard model (SM) at loop 
level, since FCNC transitions are forbidden in the SM at tree level.    
These decays induced by the FCNC are very sensitive to the
new physics beyond the SM. New physics appear in rare
decays through the Wilson coefficients which can take values different from
their SM counterpart or through the new operator structures in an effective
Hamiltonian.

Among the hadronic, leptonic and semileptonic decays, the last decay channels 
are very significant, since they are theoretically, 
more or less, clean, and they have relatively larger branching ratio. 
The semileptonic decay channels is described by the 
$b \rar s(d) \ell^+ \ell^-$ transition and they
contain many observables like forward--backward asymmetry ${\cal A}_{FB}$,
lepton polarization asymmetries, etc. Existence of these observables is very
useful and serve as a testing ground for the standard model (SM) and for
looking new physics beyond th SM. For this reason, many processes, like
$B \rar \pi(\rho) \ell^+ \ell^-$ \cite{R6602},
$B \rar \ell^+ \ell^- \gamma$ \cite{R6603},
$B \rar K \ell^+ \ell^-$ \cite{R6604} and
$B \rar K^\ast \ell^+ \ell^-$ \cite{R6605}--\cite{R6612} have been studied 
comprehensively. 

Recently, BELLE and BaBar Collaborations announced the
following results for the branching ratios of the
$B \rar K^\ast \ell^+ \ell^-$ and $B \rar K \ell^+ \ell^-$ decays:
\bea
{\cal B}(B \rar K^\ast \ell^+ \ell^-) = \left\{ \begin{array}{lc}
\left( 11.5^{+2.6}_{-2.4} \pm 0.8 \pm 0.2\right) \times
10^{-7}& \cite{R6613}~,\\ \\
\left( 0.88^{+0.33}_{-0.29} \right) \times
10^{-6}& \cite{R6614}~,\end{array} \right. \nnb
\eea
\bea
{\cal B}(B \rar K \ell^+ \ell^-) = \left\{ \begin{array}{lc}
\left( 4.8^{+1.0}_{-0.9} \pm 0.3 \pm 0.1\right) \times
10^{-7}& \cite{R6613}~,\\ \\
\left( 0.65^{+0.14}_{-0.13} \pm 0.04 \right) \times
10^{-6}& \cite{R6614}~.\end{array} \right. \nnb
\eea
Another exclusive decay which is described at inclusive level by the 
$b \rar s \ell^+ \ell^-$ transition is the baryonic 
$\Lambda_b \rar \Lambda \ell^+ \ell^-$ decay. Unlike mesonic decays, the
baryonic decays 
could maintain the helicity structure of the effective Hamiltonian for 
the $b \rar s$ transition \cite{R6615}.

Many experimentally measurable quantities such as branching ratio
\cite{R6616},
$\Lambda$ polarization and single lepton polarization have already been
studied in \cite{R6617} and \cite{R6618}, respectively. Analysis of such
quantities can be useful for more precise determination of
the SM parameters and looking for new physics beyond the SM.
It has been pointed out in \cite{R6619} that some of single lepton
polarization asymmetries may be too small to be observed and hence may not
provide sufficient number of observables in checking the structure of
effective Hamiltonian. In need of more observables, London {\it et al}
\cite{R6619} take into account polarizations of both leptons, which are
simultaneously measured, and construct maximum number of independent
observables. It should be noted that the forward--backward asymmetries
due to the double--lepton polarizations in the 
$\Lambda_b \rar \Lambda \ell^+ \ell^-$ decay, are investigated in
\cite{R6620}.

In the present work we analyze the possibility of searching for new physics
in the baryonic $\Lambda_b \rar \Lambda \ell^+ \ell^-$ decay by studying the
double--lepton polarization asymmetries, using a general form of the effective 
Hamiltonian, including all possible forms of interactions. 
Note that the dependence of polarized forward--backward asymmetry to the new 
Wilson coefficients for the meson$\rar$meson transition has been
investigated in \cite{R6611} and \cite{R6620}. In the same manner, an
analogous analysis can be carried out to investigate
the sensitivity of the double--lepton polarization asymmetries to the new
Wilson coefficients, in the case of baryon$\rar$baryon transition. 

The paper is organized as follows. In section
2, using the general, model independent form of the effective Hamiltonian,
the matrix element for the $\Lambda_b \rar \Lambda \ell^+ \ell^-$ is obtained.
In section 3 we calculate the double--lepton polarization asymmetries. 
Section 4 is devoted to the numerical analysis,
discussions and conclusions.

\section{Matrix element for the $\Lambda_b \rar \Lambda \ell^+ \ell^-$ decay}

In this section we derive the matrix element for the $\Lambda_b \rar \Lambda
\ell^+ \ell^-$ decay using the general, model independent form of the
effective Hamiltonian. At quark level, the matrix element of the 
$\Lambda_b \rar \Lambda \ell^+ \ell^-$ decay is described by the
$b \rar s \ell^+ \ell^-$ transition. The effective Hamiltonian for the
$b \rar s \ell^+ \ell^-$ transition can be written in terms of twelve model
independent four--Fermi interactions as \cite{R6606}:
\bea
\label{e6601}
{\cal M} \es \frac{G \alpha}{\sqrt{2} \pi} V_{tb}V_{ts}^\ast \Bigg\{
C_{SL} \bar s_R i \sigma_{\mu\nu} \frac{q^\nu}{q^2} b_L \bar \ell \gamma^\mu
\ell + C_{BR} \bar s_L i \sigma_{\mu\nu} \frac{q^\nu}{q^2} b_R \bar \ell
\gamma^\mu \ell + C_{LL}^{tot} \bar s_L \gamma_\mu b_L \bar \ell_L
\gamma^\mu \ell_L \nnb \\
\ar C_{LR}^{tot} \bar s_L \gamma_\mu b_L \bar \ell_R  
\gamma^\mu \ell_R + C_{RL} \bar s_R \gamma_\mu b_R \bar \ell_L
\gamma^\mu \ell_L + C_{RR} \bar s_R \gamma_\mu b_R \bar \ell_R
\gamma^\mu \ell_R \nnb \\
\ar C_{LRLR} \bar s_L b_R \bar \ell_L \ell_R +
C_{RLLR} \bar s_R b_L \bar \ell_L \ell_R +
C_{LRRL} \bar s_L b_R \bar \ell_R \ell_L +
C_{RLRL} \bar s_R b_L \bar \ell_R \ell_L \nnb \\
\ar C_T \bar s \sigma_{\mu\nu} b \bar \ell \sigma^{\mu\nu} \ell +
i C_{TE} \epsilon_{\mu\nu\alpha\beta} \bar s \sigma^{\mu\nu} b 
\bar \ell \sigma^{\alpha\beta} \ell \Bigg\}~,
\eea
where $q=P_{\Lambda_b} - P_\Lambda = p_1+p_2$ is the momentum transfer and
$C_X$ are the coefficients of the four--Fermi interactions,
$L=(1-\gamma_5)/2$ and $R=(1+\gamma_5)/2$.
The terms with coefficients $C_{SL}$ and
$C_{BR}$ describe the penguin contributions, which correspond to 
$-2 m_s C_7^{eff}$ and $-2 m_b C_7^{eff}$ in the SM, respectively. 
The next four terms in Eq. (\ref{e6601}) with coefficients
$C_{LL}^{tot},~C_{LR}^{tot},~ C_{RL}$ and $C_{RR}$
describe vector type interactions, two ($C_{LL}^{tot}$ and $C_{LR}^{tot}$)
of which contain SM contributions in the form
$C_9^{eff}-C_{10}$ and $C_9^{eff}-C_{10}$, respectively.
Thus, $C_{LL}^{tot}$ and $C_{LR}^{tot}$ can be written as 
\bea
\label{e6602}
C_{LL}^{tot} \es C_9^{eff}- C_{10} + C_{LL}~, \nnb \\
C_{LR}^{tot} \es C_9^{eff}+ C_{10} + C_{LR}~,
\eea
where $C_{LL}$ and $C_{LR}$ describe the contributions of new physics.
Additionally, Eq. (\ref{e6601}) contains four scalar type interactions 
($C_{LRLR},~C_{RLLR},~C_{LRRL}$ and $C_{RLRL}$), and two tensor type 
interactions ($C_T$ and $C_{TE}$).

The amplitude of the exclusive $\Lambda_b \rar \Lambda\ell^+ \ell^-$ decay
is obtained by calculating the matrix element of ${\cal H}_{eff}$ for the $b
\rar s \ell^+ \ell^-$ transition between initial and final
baryon states $\lla \Lambda \vel {\cal H}_{eff} \ver \Lambda_b \rra$.
It follows from Eq. (\ref{e6601}) that the matrix elements
\bea
&&\lla \Lambda \vel \bar s \gamma_\mu (1 \mp \gamma_5) b \ver \Lambda_b
\rra~,\nnb \\
&&\lla \Lambda \vel \bar s \sigma_{\mu\nu} (1 \mp \gamma_5) b \ver \Lambda_b
\rra~,\nnb \\
&&\lla \Lambda \vel \bar s (1 \mp \gamma_5) b \ver \Lambda_b \rra~.\nnb
\eea
are needed in order to calculate
the $\Lambda_b \rar \Lambda\ell^+ \ell^-$ decay amplitude.

These matrix elements parametrized in terms of the form factors are 
as follows (see \cite{R6617,R6621})
\bea
\label{e6603}
\lla \Lambda \vel \bar s \gamma_\mu b \ver \Lambda_b \rra  
\es \bar u_\Lambda \Big[ f_1 \gamma_\mu + i f_2 \sigma_{\mu\nu} q^\nu + f_3  
q_\mu \Big] u_{\Lambda_b}~,\\
\label{e6604}
\lla \Lambda \vel \bar s \gamma_\mu \gamma_5 b \ver \Lambda_b \rra
\es \bar u_\Lambda \Big[ g_1 \gamma_\mu \gamma_5 + i g_2 \sigma_{\mu\nu}
\gamma_5 q^\nu + g_3 q_\mu \gamma_5\Big] u_{\Lambda_b}~, \\
\label{e6605}
\lla \Lambda \vel \bar s \sigma_{\mu\nu} b \ver \Lambda_b \rra
\es \bar u_\Lambda \Big[ f_T \sigma_{\mu\nu} - i f_T^V \ga \gamma_\mu q^\nu -
\gamma_\nu q^\mu \dr - i f_T^S \ga P_\mu q^\nu - P_\nu q^\mu \dr \Big]
u_{\Lambda_b}~,\\
\label{e6606}
\lla \Lambda \vel \bar s \sigma_{\mu\nu} \gamma_5 b \ver \Lambda_b \rra
\es \bar u_\Lambda \Big[ g_T \sigma_{\mu\nu} - i g_T^V \ga \gamma_\mu q^\nu -
\gamma_\nu q^\mu \dr - i g_T^S \ga P_\mu q^\nu - P_\nu q^\mu \dr \Big]
\gamma_5 u_{\Lambda_b}~,
\eea
where $P = p_{\Lambda_b} + p_\Lambda$ and $q= p_{\Lambda_b} - p_\Lambda$. 

The form factors of the magnetic dipole operators are defined as 
\bea
\label{e6607}
\lla \Lambda \vel \bar s i \sigma_{\mu\nu} q^\nu  b \ver \Lambda_b \rra
\es \bar u_\Lambda \Big[ f_1^T \gamma_\mu + i f_2^T \sigma_{\mu\nu} q^\nu
+ f_3^T q_\mu \Big] u_{\Lambda_b}~,\nnb \\
\lla \Lambda \vel \bar s i \sigma_{\mu\nu}\gamma_5  q^\nu  b \ver \Lambda_b \rra
\es \bar u_\Lambda \Big[ g_1^T \gamma_\mu \gamma_5 + i g_2^T \sigma_{\mu\nu}
\gamma_5 q^\nu + g_3^T q_\mu \gamma_5\Big] u_{\Lambda_b}~.
\eea

Using the identity 
\bea
\sigma_{\mu\nu}\gamma_5 = - \frac{i}{2} \epsilon_{\mu\nu\alpha\beta}
\sigma^{\alpha\beta}~,\nnb
\eea
and Eq. (\ref{e6605}), the last expression in Eq. (\ref{e6607}) can be written as
\bea
\lla \Lambda \vel \bar s i \sigma_{\mu\nu}\gamma_5  q^\nu  b \ver \Lambda_b \rra
\es \bar u_\Lambda \Big[ f_T i \sigma_{\mu\nu} \gamma_5 q^\nu \Big]
u_{\Lambda_b}~.\nnb
\eea  
Multiplying (\ref{e6605}) and (\ref{e6606}) by $i q^\nu$ and comparing with
(\ref{e6607}), one can easily obtain the following relations between the form
factors
\bea
\label{e6608}
f_2^T \es f_T + f_T^S q^2~,\crcr
f_1^T \es \Big[ f_T^V + f_T^S \ga m_{\Lambda_b} + m_\Lambda\dr \Big] 
q^2~ = - \frac{q^2}{m_{\Lambda_b} - m_\Lambda} f_3^T~,\nnb \\
g_2^T \es g_T + g_T^S q^2~,\\
g_1^T \es \Big[ g_T^V - g_T^S \ga m_{\Lambda_b} - m_\Lambda\dr \Big]
q^2 =  \frac{q^2}{m_{\Lambda_b} + m_\Lambda} g_3^T~.\nnb
\eea 

The matrix element of the scalar $\bar s b$ and pseudoscalar
$\bar s\gamma_5 b$ operators can be obtained from (\ref{e6603}) 
and (\ref{e6604}) by multiplying both
sides to $q^\mu$ and using equation of motion. Neglecting the mass of the
strange quark, we get
\bea
\label{e6609}
\lla \Lambda \vel \bar s b \ver \Lambda_b \rra \es \frac{1}{m_b} 
\bar u_\Lambda \Big[ f_1 \ga m_{\Lambda_b} - m_\Lambda \dr + f_3 q^2
\Big] u_{\Lambda_b}~,\\
\label{e6610}
\lla \Lambda \vel \bar s \gamma_5 b \ver \Lambda_b \rra \es \frac{1}{m_b} 
\bar u_\Lambda \Big[ g_1 \ga m_{\Lambda_b} + m_\Lambda\dr \gamma_5 - g_3 q^2
\gamma_5 \Big] u_{\Lambda_b}~.
\eea

Using these definitions of the form factors, for the matrix element
of the $\Lambda_b \rar \Lambda\ell^+ \ell^-$ we get \cite{R6617,R6618}
\bea
\label{e6611}
\lefteqn{
{\cal M} = \frac{G \alpha}{4 \sqrt{2}\pi} V_{tb}V_{ts}^\ast \Bigg\{
\bar \ell \gamma^\mu \ell \, \bar u_\Lambda \Big[ A_1 \gamma_\mu (1+\gamma_5) +
B_1 \gamma_\mu (1-\gamma_5) }\nnb \\
\ar i \sigma_{\mu\nu} q^\nu \big[ A_2 (1+\gamma_5) + B_2 (1-\gamma_5) \big]
+q_\mu \big[ A_3 (1+\gamma_5) + B_3 (1-\gamma_5) \big]\Big] u_{\Lambda_b}
\nnb \\
\ar \bar \ell \gamma^\mu \gamma_5 \ell \, \bar u_\Lambda \Big[
D_1 \gamma_\mu (1+\gamma_5) + E_1 \gamma_\mu (1-\gamma_5) +
i \sigma_{\mu\nu} q^\nu \big[ D_2 (1+\gamma_5) + E_2 (1-\gamma_5) \big]
\nnb \\
\ar q_\mu \big[ D_3 (1+\gamma_5) + E_3 (1-\gamma_5) \big]\Big] u_{\Lambda_b}+
\bar \ell \ell\, \bar u_\Lambda \big(N_1 + H_1 \gamma_5\big) u_{\Lambda_b}
+\bar \ell \gamma_5 \ell \, \bar u_\Lambda \big(N_2 + H_2 \gamma_5\big) 
u_{\Lambda_b}\nnb \\
\ar 4 C_T \bar \ell \sigma^{\mu\nu}\ell \, \bar u_\Lambda \Big[ f_T 
\sigma_{\mu\nu} - i f_T^V \big( q_\nu \gamma_\mu - q_\mu \gamma_\nu \big) -
i f_T^S \big( P_\mu q_\nu - P_\nu q_\mu \big) \Big] u_{\Lambda_b}\nnb \\
\ar 4 C_{TE} \epsilon^{\mu\nu\alpha\beta} \bar \ell \sigma_{\alpha\beta}
\ell \, i \bar u_\Lambda \Big[ f_T \sigma_{\mu\nu} - 
i f_T^V \big( q_\nu \gamma_\mu - q_\mu \gamma_\nu \big) -
i f_T^S \big( P_\mu q_\nu - P_\nu q_\mu \big) \Big] u_{\Lambda_b}\Bigg\}~,
\eea
where the explicit forms of the functions $A_i,~B_i,~D_i,~E_i,~H_j$ and $N_j$
$(i=1,2,3$ and $j=1,2)$ can be written as \cite{R6617}

\bea
\label{e6612}
A_1 \es \frac{1}{q^2}\ga f_1^T-g_1^T \dr C_{SL} + \frac{1}{q^2}\ga
f_1^T+g_1^T \dr C_{BR} + \frac{1}{2}\ga f_1-g_1 \dr \ga C_{LL}^{tot} +
C_{LR}^{tot} \dr \nnb \\
\ar \frac{1}{2}\ga f_1+g_1 \dr \ga C_{RL} + C_{RR} \dr~,\nnb \\
A_2 \es A_1 \ga 1 \rar 2 \dr ~,\nnb \\
A_3 \es A_1 \ga 1 \rar 3 \dr ~,\nnb \\
B_1 \es A_1 \ga g_1 \rar - g_1;~g_1^T \rar - g_1^T \dr ~,\nnb \\
B_2 \es B_1 \ga 1 \rar 2 \dr ~,\nnb \\
B_3 \es B_1 \ga 1 \rar 3 \dr ~,\nnb \\
D_1 \es \frac{1}{2} \ga C_{RR} - C_{RL} \dr \ga f_1+g_1 \dr +
\frac{1}{2} \ga C_{LR}^{tot} - C_{LL}^{tot} \dr \ga f_1-g_1 \dr~,\nnb \\
D_2 \es D_1 \ga 1 \rar 2 \dr ~, \\
D_3 \es D_1 \ga 1 \rar 3 \dr ~,\nnb \\
E_1 \es D_1 \ga g_1 \rar - g_1 \dr ~,\nnb \\
E_2 \es E_1 \ga 1 \rar 2 \dr ~,\nnb \\
E_3 \es E_1 \ga 1 \rar 3 \dr ~,\nnb \\
N_1 \es \frac{1}{m_b} \Big( f_1 \ga m_{\Lambda_b} - m_\Lambda\dr + f_3 q^2
\Big) \Big( C_{LRLR} + C_{RLLR} + C_{LRRL} + C_{RLRL} \Big)~,\nnb \\
N_2 \es N_1 \ga C_{LRRL} \rar - C_{LRRL};~C_{RLRL} \rar - C_{RLRL} \dr~,\nnb \\
H_1 \es \frac{1}{m_b} \Big( g_1 \ga m_{\Lambda_b} + m_\Lambda\dr - g_3 q^2  
\Big) \Big( C_{LRLR} - C_{RLLR} + C_{LRRL} - C_{RLRL} \Big)~,\nnb \\
H_2 \es H_1 \ga C_{LRRL} \rar - C_{LRRL};~C_{RLRL} \rar - C_{RLRL} \dr~.\nnb
\eea

From these expressions it follows
that $\Lambda_b \rar\Lambda \ell^+\ell^-$ decay is described in terms of  
many form factors. It is shown in \cite{R6622} that HQET reduces
the number of independent form factors to two ($F_1$ and
$F_2$) irrelevant of the Dirac structure
of the corresponding operators, i.e., 
\bea
\label{e6613}
\lla \Lambda(p_\Lambda) \vel \bar s \Gamma b \ver \Lambda(p_{\Lambda_b})
\rra = \bar u_\Lambda \Big[F_1(q^2) + \not\!v F_2(q^2)\Big] \Gamma
u_{\Lambda_b}~,
\eea
where $\Gamma$ is an arbitrary Dirac structure and
$v^\mu=p_{\Lambda_b}^\mu/m_{\Lambda_b}$ is the four--velocity of
$\Lambda_b$. Comparing the general form of the form factors given in Eqs.
(\ref{e6603})--(\ref{e6610}) with (\ref{e6613}), one can
easily obtain the following relations among them (see also
\cite{R6617,R6618,R6621})
\bea
\label{e6614}
g_1 \es f_1 = f_2^T= g_2^T = F_1 + \sqrt{\hat{r}_\Lambda} F_2~, \nnb \\
g_2 \es f_2 = g_3 = f_3 = g_T^V = f_T^V = \frac{F_2}{m_{\Lambda_b}}~,\nnb \\
g_T^S \es f_T^S = 0 ~,\nnb \\
g_1^T \es f_1^T = \frac{F_2}{m_{\Lambda_b}} q^2~,\nnb \\
g_3^T \es \frac{F_2}{m_{\Lambda_b}} \ga m_{\Lambda_b} + m_\Lambda \dr~,\nnb \\
f_3^T \es - \frac{F_2}{m_{\Lambda_b}} \ga m_{\Lambda_b} - m_\Lambda \dr~,
\eea
where $\hat{r}_\Lambda=m_\Lambda^2/m_{\Lambda_b}^2$.

From Eq. (\ref{e6611}), we get for the unpolarized decay width
\bea
\label{e6615}
\ga \frac{d \Gamma}{d\hat{s}}\dr_0 = \frac{G^2 \alpha^2}{8192 \pi^5}
\vel V_{tb} V_{ts}^\ast \ver^2 \lambda^{1/2}(1,\hat{r}_\Lambda,\hat{s}) v
\Bigg[{\cal T}_0(\hat{s}) +\frac{1}{3} {\cal T}_2(\hat{s}) \Bigg]~,
\eea
where 
$\lambda(1,\hat{r}_\Lambda,\hat{s}) = 1 + \hat{r}_\Lambda^2 + \hat{s}^2 - 
2 \hat{r}_\Lambda - 2 \hat{s} - 2 \hat{r}_\Lambda\hat{s}$
is the triangle function, $\hat{s} = q^2/m_{\Lambda_b}^2$ and
$v=\sqrt{1-4 \hat{m}_\ell^2/\hat{s}}$ is the lepton
velocity, with $\hat{m}_\ell = m_\ell/m_{\Lambda_b}$. 
The explicit expressions for ${\cal T}_0$ and ${\cal T}_2$ can be 
found in \cite{R6618}. In the next section, we present the expressions for 
the double--lepton polarization asymmetries.

\section{Double--lepton polarization asymmetries in the 
$\Lambda_b \rar\Lambda \ell^+\ell^-$ decay}

In the present section we calculate the double--lepton polarization asymmetries,
i.e., when polarizations of both leptons are considered.
In order to calculate the double lepton polarization asymmetries, 
we define the following orthogonal unit vectors
$s_i^{\pm\mu}$ in the rest frame of $\ell^\pm$ 
($i=L,T$ or $N$, stand for longitudinal, transversal or
normal polarizations, respectively.

\bea
\label{e6616}   
s^{-\mu}_L \es \ga 0,\vec{e}_L^{\,-}\dr =
\ga 0,\frac{\vec{p}_-}{\vel\vec{p}_- \ver}\dr~, \nnb \\
s^{-\mu}_N \es \ga 0,\vec{e}_N^{\,-}\dr = \ga 0,\frac{\vec{p}_\Lambda\times
\vec{p}_-}{\vel \vec{p}_\Lambda\times \vec{p}_- \ver}\dr~, \nnb \\
s^{-\mu}_T \es \ga 0,\vec{e}_T^{\,-}\dr = \ga 0,\vec{e}_N^{\,-}
\times \vec{e}_L^{\,-} \dr~, \nnb \\
s^{+\mu}_L \es \ga 0,\vec{e}_L^{\,+}\dr =
\ga 0,\frac{\vec{p}_+}{\vel\vec{p}_+ \ver}\dr~, \nnb \\
s^{+\mu}_N \es \ga 0,\vec{e}_N^{\,+}\dr = \ga 0,\frac{\vec{p}_\Lambda\times
\vec{p}_+}{\vel \vec{p}_\Lambda\times \vec{p}_+ \ver}\dr~, \nnb \\
s^{+\mu}_T \es \ga 0,\vec{e}_T^{\,+}\dr = \ga 0,\vec{e}_N^{\,+}
\times \vec{e}_L^{\,+}\dr~,
\eea
where $\vec{p}_\mp$ and $\vec{p}_\Lambda$ are the three--momenta of the
leptons $\ell^\mp$ and $\Lambda$ baryon in the
center of mass frame (CM) of $\ell^- \,\ell^+$ system, respectively.
Transformation of unit vectors from the rest frame of the leptons to CM
frame of leptons can be accomplished by the Lorentz boost. Boosting of the
longitudinal unit vectors $s_L^{\pm\mu}$ yields
\bea
\label{e6617}
\ga s^{\mp\mu}_L \dr_{CM} \es \ga \frac{\vel\vec{p}_\mp \ver}{m_\ell}~,
\frac{E_\ell \vec{p}_\mp}{m_\ell \vel\vec{p}_\mp \ver}\dr~,
\eea
where $\vec{p}_+ = - \vec{p}_-$, $E_\ell$ and $m_\ell$ are the energy and mass
of leptons in the CM frame, respectively.
The remaining two unit vectors $s_N^{\pm\mu}$, $s_T^{\pm\mu}$ are unchanged
under Lorentz boost.

having obtained the above--expressions, we now define the double--polarization 
asymmetries as follows \cite{R6619}:

\bea                                                                  
\label{e6618}
P_{ij}(q^2) \es
\frac{ 
\Big( \ds \frac{d\Gamma(\vec{s}^{\,-}_i,\vec{s}^{\,+}_j)}{dq^2} -
      \ds \frac{d\Gamma(-\vec{s}^{\,-}_i,\vec{s}^{\,+}_j)}{dq^2} \Big) -
\Big( \ds \frac{d\Gamma(\vec{s}^{\,-}_i,-\vec{s}^{\,+}_j)}{dq^2} -      
      \ds \frac{d\Gamma(-\vec{s}^{\,-}_i,-\vec{s}^{\,+}_j)}{dq^2} \Big) 
     }
     {    
\Big( \ds \frac{d\Gamma(\vec{s}^{\,-}_i,\vec{s}^{\,+}_j)}{dq^2} +      
      \ds \frac{d\Gamma(-\vec{s}^{\,-}_i,\vec{s}^{\,+}_j)}{dq^2} \Big) +
\Big( \ds \frac{d\Gamma(\vec{s}^{\,-}_i,-\vec{s}^{\,+}_j)}{dq^2} +      
      \ds \frac{d\Gamma(-\vec{s}^{\,-}_i,-\vec{s}^{\,+}_j)}{dq^2} \Big)
     }~,
\eea
where, the first subindex $i$ represents lepton and the second one
antilepton. Using this definition of $P_{ij}$, nine double--lepton
polarization asymmetries are calculated. Their expressions are

\bea
\label{e6619}
P_{LL} \es \frac{8 m_{\Lambda_b}^4}{3 \Delta}
\mbox{\rm Re} \Bigg\{
%1
32 m_{\Lambda_b}^2 \hat{m}_\ell \lambda
\Big(1+\sqrt{\hat{r}_\Lambda}\Big)
(A_1+B_1) C_{T}^\ast f_T^{S\ast} \nnb \\
%2
\ek 12 \Big(1+\sqrt{\hat{r}_\Lambda}\Big)
\Big(1-2\sqrt{\hat{r}_\Lambda}+\hat{r}_\Lambda - \hat{s}\Big)
\Big\{ \hat{m}_\ell (D_1-E_1) H_2^\ast \nnb \\
\ar 8 m_{\Lambda_b}^2 \hat{m}_\ell \hat{s}
(A_2+B_2) C_{T} f_T^{V\ast} +
32 m_{\Lambda_b} \hat{s} \Big[
\vel C_{T} \ver^2 ( 2 v^2 -1) +
4 v^2 \vel C_{TE} \ver^2 \Big] f_T f_T^{V\ast} \Big\} \nnb \\
%3
\ar 12 \hat{m}_\ell \Big(1-\sqrt{\hat{r}_\Lambda}\Big)
\Big(1+2\sqrt{\hat{r}_\Lambda}+\hat{r}_\Lambda - \hat{s}\Big)
(D_1+E_1) F_2^\ast \nnb \\
%4
\ek 32 m_{\Lambda_b}^3 \hat{m}_\ell \lambda \hat{s} 
(A_2+B_2) C_{T}^\ast f_T^{S\ast} \nnb \\
%5
\ar 3 \hat{s} \Big(1+2\sqrt{\hat{r}_\Lambda}+\hat{r}_\Lambda - 
\hat{s}\Big) \Big[ v^2 \vel F_1 \ver^2 + \vel F_2 \ver^2 + 
4 m_{\Lambda_b}  \hat{m}_\ell (D_3+E_3) F_2^\ast \Big] \nnb \\
%6
\ek 12 m_{\Lambda_b} \sqrt{\hat{r}_\Lambda}
(1-\hat{r}_\Lambda+\hat{s})
\Big[ \hat{s} (1+v^2) (A_1 A_2^\ast + B_1 B_2^\ast)  - 
4 \hat{m}_\ell^2 (D_1 D_3^\ast + E_1 E_3^\ast) \Big] \nnb \\
%7
\ar 3 \hat{s} \Big(1-2\sqrt{\hat{r}_\Lambda}+
\hat{r}_\Lambda - \hat{s}\Big) \Big[4 m_{\Lambda_b} \hat{m}_\ell 
(D_3-E_3) H_2^\ast + v^2 \vel H_1 \ver^2 + \vel H_2 \ver^2 \Big] \nnb \\
%8
\ar 12 m_{\Lambda_b} (1-\hat{r}_\Lambda-\hat{s})
\Big[ \hat{s} (1+v^2) (A_1 B_2^\ast + A_2 B_1^\ast) + 
4 \hat{m}_\ell^2 (D_1 E_3^\ast + D_3 E_1^\ast) \Big] \nnb \\
%9
\ar 24 \sqrt{\hat{r}_\Lambda} \hat{s} (1+v^2) 
\Big( A_1 B_1^\ast + D_1 E_1^\ast +
m_{\Lambda_b}^2 \hat{s} A_2 B_2^\ast \Big) \nnb \\
%10
\ar 48 m_{\Lambda_b}^2 \hat{m}_\ell^2 \hat{s} (1+\hat{r}_\Lambda-\hat{s})
D_3 E_3^\ast \nnb \\
%11
\ar 16 m_{\Lambda_b} \Big[ (1-\hat{r}_\Lambda)^2 + 
\hat{s} \Big(1-6 \sqrt{\hat{r}_\Lambda} + \hat{r}_\Lambda \Big) -
2 \hat{s}^2 \Big] \Big[ 2 \hat{m}_\ell 
(A_1+B_1) C_{T}^\ast f_T^{V\ast} \nnb \\
\ek 32 m_{\Lambda_b} \hat{m}_\ell^2 \Big(\vel C_{T} \ver^2 +  
2 \vel C_{TE} \ver^2\Big) \vel f_T^{V\ast} \ver^2\Big) +
4 m_{\Lambda_b} \hat{s} \Big( \vel C_{T} \ver^2 +  
4 \vel C_{TE} \ver^2\Big) \vel f_T^{V\ast} \ver^2\Big) \Big] \nnb \\
%12
\ek 2 (1+v^2)
\Big[ 1+\hat{r}_\Lambda^2 - 
\hat{r}_\Lambda (2-\hat{s}) +\hat{s} (1-2 \hat{s}) \Big]
\Big(\vel A_1 \ver^2 + \vel B_1 \ver^2 \Big) \nnb \\
%13
\ek 2 \Big[
(5 v^2 - 3) (1-\hat{r}_\Lambda)^2 +   
4 \hat{m}_\ell^2 (1+\hat{r}_\Lambda) +
2 \hat{s} (1+8 \hat{m}_\ell^2 + \hat{r}_\Lambda)
- 4 \hat{s}^2 \Big] \Big( \vel D_1 \ver^2 + \vel E_1 \ver^2 \Big) \nnb \\
%14
\ek 2 m_{\Lambda_b}^2 (1+v^2) \hat{s}
\Big[2 + 2 \hat{r}_\Lambda^2 -\hat{s}(1 +\hat{s}) -
\hat{r}_\Lambda (4 + \hat{s})\Big] \big(
\vel A_2 \ver^2 + \vel B_2 \ver^2 \Big) \nnb \\
%15
\ar 64 m_{\Lambda_b}^3 \lambda \hat{s}\Big[  
2 \Big(1+\sqrt{\hat{r}_\Lambda}\Big) f_T^{S} f_T^{V\ast} +
m_{\Lambda_b} \Big(1+2 \sqrt{\hat{r}_\Lambda} + \hat{r}_\Lambda
-\hat{s}\Big)
\vel f_T^S \ver^2 \Big]
\Big[ (2 v^2-1) \vel C_{T} \ver^2 + 4 v^2 
\vel C_{TE} \ver^2 \Big] \nnb \\
%16
\ek 4 m_{\Lambda_b}^2 \hat{s} v^2 \Big[
2 (1 + \hat{r}_\Lambda^2) - \hat{s} (1+\hat{s}) - 
\hat{r}_\Lambda (4+\hat{s})\Big] \Big(
\vel D_2 \ver^2 + \vel E_2 \ver^2 \Big) \nnb \\
%17
\ar 24 m_{\Lambda_b} \hat{s} (1-\hat{r}_\Lambda-\hat{s}) v^2
\Big( D_1 E_2^\ast + D_2 E_1^\ast \Big) \nnb \\
%18
\ek 24 m_{\Lambda_b} \sqrt{\hat{r}_\Lambda} \hat{s}
(1-\hat{r}_\Lambda+\hat{s}) v^2
\Big( D_1 D_2^\ast + E_1 E_2^\ast \Big) \nnb \\
%19
\ar 48 m_{\Lambda_b}^2 \sqrt{\hat{r}_\Lambda} \hat{s}
\Big( \hat{s} v^2 D_2 E_2^\ast + 
2 \hat{m}_\ell^2 D_3 E_3^\ast \Big) \nnb \\
%20
\ek 128 m_{\Lambda_b}^2 \lambda \hat{s}                
\Big[ (2 v^2 -1) \vel C_{T} \ver^2 + 4 v^2 \vel C_{TE} \ver^2 \Big] 
f_T f_T^{S\ast} \nnb \\
%21
\ar 32 m_{\Lambda_b} \hat{m}_\ell
\Big\{ \Big[ 2 (1-\hat{r}_\Lambda)^2 - \hat{s}(1+\hat{r}_\Lambda)
- \hat{s}^2 \Big] (C_{T} - 2 C_{TE})  
- 6 \sqrt{\hat{r}_\Lambda} \hat{s} (C_{T}+2 C_{TE})
\Big\} A_2^\ast f_T  \nnb \\
%22
\ar 32 m_{\Lambda_b} \hat{m}_\ell
\Big\{ \Big[ 2 (1-\hat{r}_\Lambda)^2 - \hat{s}(1+\hat{r}_\Lambda)
- \hat{s}^2 \Big] (C_{T} + 2 C_{TE})
- 6 \sqrt{\hat{r}_\Lambda} \hat{s} (C_{T}-2 C_{TE})
\Big\} B_2^\ast f_T \nnb \\
%23
\ek 96 \hat{m}_\ell
\Big[ \Big(1 + \sqrt{\hat{r}_\Lambda}\Big) \Big(1 - 2 \sqrt{\hat{r}_\Lambda} 
+ \hat{r}_\Lambda - \hat{s}\Big) C_{T} +   
2 \Big(1 - \sqrt{\hat{r}_\Lambda}\Big) \Big(1 + 2 \sqrt{\hat{r}_\Lambda} 
+ \hat{r}_\Lambda - \hat{s}\Big) C_{TE} \Big] A_1^\ast f_T \nnb \\
%24
\ek 96 \hat{m}_\ell
\Big[ \Big(1 + \sqrt{\hat{r}_\Lambda}\Big) \Big(1 - 2 \sqrt{\hat{r}_\Lambda}
+ \hat{r}_\Lambda - \hat{s}\Big) C_{T} -
2 \Big(1 - \sqrt{\hat{r}_\Lambda}\Big) \Big(1 + 2 \sqrt{\hat{r}_\Lambda}
+ \hat{r}_\Lambda - \hat{s}\Big) C_{TE} \Big] B_1^\ast f_T \nnb \\
%25-1
\ar 128 \Big[ (3 v^2-1) (1-\hat{r}_\Lambda)^2 +
6 \hat{m}_\ell^2 \Big( 1 + 2 \sqrt{\hat{r}_\Lambda} + \hat{r}_\Lambda
+ \hat{s} \Big) - \hat{s} (1+\hat{r}_\Lambda+\hat{s}) \Big]
\vel C_{T} \ver^2 \vel f_T \ver^2 \nnb \\
%25-2                                                                
\ar 512 \Big[ (3 v^2-1) (1-\hat{r}_\Lambda)^2 +
6 \hat{m}_\ell^2 \Big( 1 - 2 \sqrt{\hat{r}_\Lambda} + \hat{r}_\Lambda
+ \hat{s} \Big) - \hat{s} (1+\hat{r}_\Lambda+\hat{s}) \Big]
\vel C_{TE} \ver^2 \vel f_T \ver^2
\Bigg\}~, \\ \nnb \\
\label{e6620}
P_{LN} \es \frac{4 \pi m_{\Lambda_b}^4 \sqrt{\lambda}}{\Delta \sqrt{\hat{s}}} 
\mbox{\rm Im} \Bigg\{
%1
4 \hat{m}_\ell  
(1-\hat{r}_\Lambda) 
(A_1^\ast D_1 + B_1^\ast E_1) \nnb \\
%2
\ar 32 m_{\Lambda_b}^2 \hat{m}_\ell^2 
 \Big(1-\sqrt{\hat{r}_\Lambda} \Big)
\Big(1+2\sqrt{\hat{r}_\Lambda}+\hat{r}_\Lambda-\hat{s} \Big)
(D_1^\ast + E_1^\ast ) C_{T} f_T^S \nnb \\
%3
\ar 16 \hat{m}_\ell  
 \hat{s} (F_2 C_{T}^\ast f_T^\ast - 
2 H_2 C_{TE}^\ast f_T^\ast) \nnb \\
%4
\ar 4 m_{\Lambda_b} \hat{m}_\ell  
 \hat{s} (A_1^\ast E_3 - A_2^\ast E_1 + B_1^\ast D_3
-B_2^\ast D_1) \nnb \\
%5
\ar  \hat{s}
\Big(1+\sqrt{\hat{r}_\Lambda} \Big) 
\Big\{(A_1^\ast+B_1^\ast) F_2 - 16 m_{\Lambda_b} \hat{m}_\ell
\Big[ F_2 + 2 m_{\Lambda_b} \hat{m}_\ell (D_3 + E_3) \Big]  
C_{T}^\ast f_T^{V\ast} \Big\} \nnb \\
%6
\ar 32 m_{\Lambda_b} \hat{m}_\ell^2 
 \hat{s}
\Big[ (D_3+E_3) C_{T}^\ast f_T^\ast - 2 (D_3-E_3) 
C_{TE}^\ast f_T^\ast \Big] \nnb \\
%7
\ar  \hat{s} \Big(1-\sqrt{\hat{r}_\Lambda} \Big)
(A_1-B_1) H_2^\ast \nnb \\
%8
\ar  \hat{s}
\Big[ \Big(1-\sqrt{\hat{r}_\Lambda} \Big) (D_1+E_1) C_{T}^\ast f_T^\ast +
2 \Big(1+\sqrt{\hat{r}_\Lambda} \Big) (D_1-E_1) 
C_{TE}^\ast f_T^\ast \Big] \nnb \\
%9
\ar 4 m_{\Lambda_b} \hat{m}_\ell
 \sqrt{\hat{r}_\Lambda} \hat{s}
(A_1^\ast D_3 + A_2^\ast D_1 +B_1^\ast E_3 + B_2^\ast E_1) \nnb \\
%10
\ek 16 m_{\Lambda_b}^2 \hat{m}_\ell
 \hat{s}
\Big(1+2 \sqrt{\hat{r}_\Lambda}+\hat{r}_\Lambda -\hat{s} \Big)
\Big\{\Big[F_2 + 2 m_{\Lambda_b} \hat{m}_\ell (D_3+E_3) \Big] 
C_{T}^\ast f_T^{S\ast} \Big\} \nnb \\
%11
\ek m_{\Lambda_b}  \hat{s}^2
\Big[ B_2^\ast (F_2 - H_2 + 4 m_{\Lambda_b} \hat{m}_\ell E_3)
+ A_2^\ast (F_2 + H_2 + 4 m_{\Lambda_b} \hat{m}_\ell D_3) -
v^2 D_2^\ast F_1 \Big] \nnb \\
%12
\ar  \hat{s} v^2
\Big\{(D_1 + E_1) F_1^\ast - (D_1 - E_1) H_1^\ast -
\sqrt{\hat{r}_\Lambda} \Big[ E_1^\ast (F_1-H_1) + D_1^\ast (F_1+H_1)
\Big] \Big\} \nnb \\
%13
\ar 8  \hat{s} v^2
\Big\{(A_1 - B_1) C_{T}^\ast f_T^\ast + 2 (A_1 + B_1) 
C_{TE}^\ast f_T^\ast \nnb \\
\ek
\sqrt{\hat{r}_\Lambda} \Big[ A_1^\ast (C_T-2 C_{TE}) f_T + 
B_1^\ast (C_T+2 C_{TE})f_T \Big] \Big\} \nnb \\
%14
\ar 8 m_{\Lambda_b}  \hat{s}    
(1 - \hat{r}_\Lambda) v^2
\Big[ (A_2-B_2-D_2-E_2) C_{T}^\ast f_T^\ast - 2 (A_2 +B_2 - D_2+E_2)
C_{TE}^\ast f_T^\ast \Big] \nnb \\
%15
\ek m_{\Lambda_b}  \hat{s}^2 v^2
\Big[ 8 (A_1-B_1) C_{T}^\ast f_T^{V\ast} - (F_1-H_1) E_2^\ast 
+ D_2 H_1^\ast \Big] \nnb \\
%16
\ek 8 m_{\Lambda_b}^2  \hat{s}^2  
\Big(1 - \sqrt{\hat{r}_\Lambda} \Big) v^2
\Big[ (A_2-B_2) C_{T}^\ast f_T^{V\ast} + 
2 (D_2-E_2) C_{TE}^\ast f_T^{V\ast} \Big] \nnb \\
%17
\ek 16 m_{\Lambda_b} 
\Big[ 2 \hat{m}_\ell^2 (1-\hat{r}_\Lambda) 
(D_1+E_1) C_{T}^\ast f_T^{V\ast} +
\hat{s}^2 v^2 (D_1-E_1) C_{TE}^\ast f_T^{V\ast}\Big] 
\Bigg\}~, \\ \nnb \\
\label{e6621}
P_{NL} \es - \frac{4 \pi m_{\Lambda_b}^4 \sqrt{\lambda}}{\Delta \sqrt{\hat{s}}} 
\mbox{\rm Im} \Bigg\{
%1
4\hat{m}_\ell  
(1-\hat{r}_\Lambda) 
(A_1^\ast D_1 + B_1^\ast E_1) \nnb \\
%2
\ar 32 m_{\Lambda_b}^2 \hat{m}_\ell^2
 \Big(1-\sqrt{\hat{r}_\Lambda} \Big)
\Big(1+2\sqrt{\hat{r}_\Lambda}+\hat{r}_\Lambda-\hat{s} \Big)
(D_1^\ast + E_1^\ast ) C_{T} f_T^S \nnb \\
%3
\ar 16 \hat{m}_\ell  
 \hat{s} (F_2 C_{T}^\ast f_T^\ast -
2 H_2 C_{TE}^\ast f_T^\ast) \nnb \\
%4
\ar 4 m_{\Lambda_b} \hat{m}_\ell  
 \hat{s} (A_1^\ast E_3 - A_2^\ast E_1 + B_1^\ast D_3
-B_2^\ast D_1) \nnb \\
%5
\ar   \hat{s}
\Big(1+\sqrt{\hat{r}_\Lambda} \Big) 
\Big\{(A_1^\ast+B_1^\ast) F_2 - 16 m_{\Lambda_b} \hat{m}_\ell
\Big[ F_2 + 2 m_{\Lambda_b} \hat{m}_\ell (D_3 + E_3) \Big]
C_{T}^\ast f_T^{V\ast} \Big\} \nnb \\
%6
\ar 32  m_{\Lambda_b} \hat{m}_\ell^2 
 \hat{s}
\Big[ (D_3+E_3) C_{T}^\ast f_T^\ast - 2 (D_3-E_3)
C_{TE}^\ast f_T^\ast \Big] \nnb \\
%7
\ar  \hat{s} \Big(1-\sqrt{\hat{r}_\Lambda} \Big)
(A_1-B_1) H_2^\ast \nnb \\
%8
\ar  \hat{s}
\Big[ \Big(1-\sqrt{\hat{r}_\Lambda} \Big) (D_1+E_1) C_{T}^\ast f_T^\ast +
2 \Big(1+\sqrt{\hat{r}_\Lambda} \Big) (D_1-E_1) 
C_{TE}^\ast f_T^\ast \Big] \nnb \\
%9
\ar 4 m_{\Lambda_b} \hat{m}_\ell
 \sqrt{\hat{r}_\Lambda} \hat{s}
(A_1^\ast D_3 + A_2^\ast D_1 +B_1^\ast E_3 + B_2^\ast E_1) \nnb \\
%10
\ek 4  m_{\Lambda_b} \hat{m}_\ell
 \sqrt{\hat{r}_\Lambda} \hat{s}
(A_1^\ast D_3 + A_2^\ast D_1 +B_1^\ast E_3 + B_2^\ast E_1) \nnb \\
%11
\ek m_{\Lambda_b}  \hat{s}^2
\Big[ B_2^\ast (F_2 - H_2 + 4 m_{\Lambda_b} \hat{m}_\ell E_3)
+ A_2^\ast (F_2 + H_2 + 4 m_{\Lambda_b} \hat{m}_\ell D_3) -
v^2 D_2^\ast F_1 \Big] \nnb \\
%12
\ek  \hat{s} v^2
\Big\{(D_1 + E_1) F_1^\ast - (D_1 - E_1) H_1^\ast \nnb \\
\ek
\sqrt{\hat{r}_\Lambda} \Big[ E_1^\ast (F_1-H_1) + D_1^\ast (F_1+H_1)
\Big] \Big\} \nnb \\
%13
\ek 8  \hat{s} v^2
\Big\{(A_1 - B_1) C_{T}^\ast f_T^\ast + 2 (A_1 + B_1) 
C_{TE}^\ast f_T^\ast \nnb \\
\ek
\sqrt{\hat{r}_\Lambda} \Big[ A_1^\ast (C_T-2 C_{TE}) f_T + 
B_1^\ast (C_T+2 C_{TE})f_T \Big] \Big\} \nnb \\
%14
\ek 8 m_{\Lambda_b}  \hat{s}    
(1 - \hat{r}_\Lambda) v^2
\Big[ (A_2-B_2+D_2+E_2) C_{T}^\ast f_T^\ast - 2 (A_2 +B_2 +D_2-E_2)
C_{TE}^\ast f_T^\ast \Big] \nnb \\
%15
\ar m_{\Lambda_b}  \hat{s}^2 v^2
\Big[ 8 (A_1-B_1) C_{T}^\ast f_T^{V\ast} - (F_1-H_1) E_2^\ast 
+ D_2 H_1^\ast \Big] \nnb \\
%16
\ar 8 m_{\Lambda_b}^2  \hat{s}^2  
\Big(1 - \sqrt{\hat{r}_\Lambda} \Big) v^2
\Big[ (A_2-B_2) C_{T}^\ast f_T^{V\ast} - 
2 (D_2-E_2) C_{TE}^\ast f_T^{V\ast} \Big] \nnb \\
%17
\ek 16 m_{\Lambda_b} 
\Big[ 2 \hat{m}_\ell^2 (1-\hat{r}_\Lambda)
(D_1+E_1) C_{T}^\ast f_T^{V\ast} +
\hat{s}^2 v^2 (D_1-E_1) C_{TE}^\ast f_T^{V\ast}\Big]
\Bigg\}~, \\ \nnb \\
\label{e6622}
P_{LT} \es \frac{4 \pi m_{\Lambda_b}^4 \sqrt{\lambda} v}{\Delta \sqrt{\hat{s}}} 
\mbox{\rm Re} \Bigg\{
%1
4 \hat{m}_\ell 
(1-\hat{r}_\Lambda) \Big[ \vel D_1 \ver^2 + \vel E_1 \ver^2
- 32 \Big( \vel C_{T} \ver^2 + 4 \vel C_{TE} \ver^2 \Big) 
\vel f_T \ver^2 \Big] \nnb \\
%2
\ek 4 \hat{m}_\ell
\hat{s} \Big(A_1 D_1^\ast - B_1 E_1^\ast + 
4 F_1^\ast  C_{T} f_T \Big) \nnb \\
%3
\ar 4 \hat{m}_\ell
\hat{s} \Big\{ 8 H_1 C_{TE}^\ast f_T^\ast - m_{\Lambda_b}                   
\Big[ B_1 D_2^\ast + (A_2 + D_2 -D_3) E_1^\ast \nnb \\
\ek  A_1 E_2^\ast
-(B_2-E_2+E_3) D_1^\ast \Big] \Big\} \nnb \\
%4
\ar 
\hat{s} \Big(1 - \sqrt{\hat{r}_\Lambda} \Big) 
\Big[ (A_1 - B_1) H_1^\ast - (D_1-E_1) H_2^\ast \nnb \\
\ar 128 m_{\Lambda_b} \hat{m}_\ell \Big( 4 \vel C_{TE} \ver^2 + 
\vel C_{T} \ver^2 \Big) f_T f_T^{V\ast}\Big] \nnb \\
%5
\ek 
\hat{s} \Big(1 + \sqrt{\hat{r}_\Lambda} \Big) 
\Big[ (A_1 + B_1) F_1^\ast - (D_1+E_1) F_2^\ast \nnb \\
\ek 16 m_{\Lambda_b} \hat{m}_\ell F_1 C_{T}^\ast f_T^{V\ast} \Big] \nnb \\
%6
\ek 8 
\hat{s}                                      
\Big\{ \Big[(A_1^\ast + D_1^\ast) - 
\sqrt{\hat{r}_\Lambda} (B_1^\ast + E_1^\ast)\Big]
(C_{T} + 2 C_{TE}) f_T \nnb \\
\ar \Big[(B_1^\ast - E_1^\ast) - 
\sqrt{\hat{r}_\Lambda} (A_1^\ast - D_1^\ast)\Big]
(C_{T} - 2 C_{TE}) f_T \Big\} \nnb \\
%7
\ar 4 m_{\Lambda_b} 
\hat{s} (1-\hat{r}_\Lambda)
\Big\{ 2 \Big[A_2^\ast (C_{T} - 2 C_{TE}) f_T
+ B_2^\ast (C_{T} + 2 C_{TE}) f_T \nnb \\
\ek D_2^\ast (C_{T} - 2 C_{TE}) f_T +
E_2^\ast (C_{T} + 2 C_{TE}) f_T \Big]
+ m_{\Lambda_b} \hat{m}_\ell (A_2 D_2^\ast - B_2 E_2^\ast) \Big\}\nnb \\
%8
\ar 4 m_{\Lambda_b} \hat{m}_\ell 
 \sqrt{\hat{r}_\Lambda} \hat{s}
\Big[ A_1 D_2^\ast + (A_2 + D_2 +D_3) D_1^\ast - B_1 E_2^\ast -
(B_2 - E_2 - E_3) E_1^\ast \Big] \nnb \\ 
%9
\ar m_{\Lambda_b}             
 \hat{s}^2
\Big\{ 
8 \Big[2 (A_1 - B_1) C_{TE}^\ast + (D_1 - E_1) C_{T}^\ast\Big]
f_T^{V\ast} + (A_2-B_2) H_1^\ast \nnb \\
\ar (A_2 + B_2) F_1^\ast - (D_2-E_2)
H_2^\ast - (D_2+E_2) F_2^\ast - 
4 m_{\Lambda_b} \hat{m}_\ell (D_2 D_3^\ast + E_2 E_3^\ast )
\Big\} \nnb \\
%10
\ar 8 m_{\Lambda_b}^2
 \hat{s}^2 \Big(1-\sqrt{\hat{r}_\Lambda}\Big)
\Big[ 2 (A_2-B_2) C_{TE}^\ast + (D_2-E_2) C_{T}^\ast
\Big] f_T^{V\ast} \nnb \\
%11
\ar 16 m_{\Lambda_b}^2 \hat{m}_\ell   
 \hat{s} \Big(1+2 \sqrt{\hat{r}_\Lambda}+\hat{r}_\Lambda
-\hat{s}\Big) F_1^\ast C_{T} f_T^S
\Bigg\}~, \\ \nnb \\
\label{e6623}
P_{TL} \es \frac{4 \pi m_{\Lambda_b}^4 \sqrt{\lambda} v}{\Delta \sqrt{\hat{s}}} 
\mbox{\rm Re} \Bigg\{
%1
4 \hat{m}_\ell 
(1-\hat{r}_\Lambda) \Big[ \vel D_1 \ver^2 + \vel E_1 \ver^2
- 32 \Big( \vel C_{T} \ver^2 + 4 \vel C_{TE} \ver^2 \Big)
\vel f_T \ver^2 \Big] \nnb \\
%2
\ar 4 \hat{m}_\ell
\hat{s} \Big(A_1 D_1^\ast - B_1 E_1^\ast + 
4 F_1^\ast  C_{T} f_T \Big) \nnb \\
%3
\ek 4 \hat{m}_\ell
\hat{s} \Big\{ 8 H_1 C_{TE}^\ast f_T^\ast - m_{\Lambda_b}                   
\Big[ B_1 D_2^\ast + (A_2 - D_2 + D_3) E_1^\ast \nnb \\
\ek  A_1 E_2^\ast
- (B_2+E_2-E_3) D_1^\ast \Big] \Big\} \nnb \\
%4
\ek 
\hat{s} \Big(1 - \sqrt{\hat{r}_\Lambda} \Big)
\Big[ (A_1 - B_1) H_1^\ast + (D_1-E_1) H_2^\ast \nnb \\
\ek 128 m_{\Lambda_b} \hat{m}_\ell \Big( 4 \vel C_{TE} \ver^2 +
\vel C_{T} \ver^2 \Big) f_T f_T^{V\ast}\Big] \nnb \\
%5
\ar 
\hat{s} \Big(1 + \sqrt{\hat{r}_\Lambda} \Big)
\Big[ (A_1 + B_1) F_1^\ast + (D_1+E_1) F_2^\ast \nnb \\
\ek 16 m_{\Lambda_b} \hat{m}_\ell F_1 C_{T}^\ast f_T^{V\ast} \Big] \nnb \\
%6    
\ek 8 
\hat{s}
\Big\{ \Big[(A_1^\ast - D_1^\ast) - 
\sqrt{\hat{r}_\Lambda} (B_1^\ast - E_1^\ast)\Big]
(C_{T} + 2 C_{TE}) f_T \nnb \\
\ar \Big[(B_1^\ast + E_1^\ast) - 
\sqrt{\hat{r}_\Lambda} (A_1^\ast + D_1^\ast)\Big]
(C_{T} - 2 C_{TE}) f_T \Big\} \nnb \\
%7
\ar 4 m_{\Lambda_b} 
\hat{s} (1-\hat{r}_\Lambda)
\Big\{ 2 \Big[A_2^\ast (C_{T} - 2 C_{TE}) f_T +
B_2^\ast (C_{T} + 2 C_{TE}) f_T \nnb \\
\ar D_2^\ast (C_{T} - 2 C_{TE}) f_T -
E_2^\ast (C_{T} + 2 C_{TE}) f_T \Big] -
m_{\Lambda_b} \hat{m}_\ell (A_2 D_2^\ast - B_2 E_2^\ast) \Big\} \nnb \\
%8
\ek 4 m_{\Lambda_b} \hat{m}_\ell 
 \sqrt{\hat{r}_\Lambda} \hat{s}
\Big[ A_1 D_2^\ast + (A_2 - D_2 - D_3) D_1^\ast - B_1 E_2^\ast -
(B_2 + E_2 + E_3) E_1^\ast \Big] \nnb \\
%9
\ar m_{\Lambda_b}                       
 \hat{s}^2
\Big\{
8 \Big[2 (A_1 - B_1) C_{TE}^\ast - (D_1 - E_1) C_{T}^\ast\Big]
f_T^{V\ast} - (A_2-B_2) H_1^\ast \nnb \\
\ek (A_2 + B_2) F_1^\ast - (D_2-E_2)
H_2^\ast - (D_2+E_2) F_2^\ast -
4 m_{\Lambda_b} \hat{m}_\ell (D_2 D_3^\ast + E_2 E_3^\ast )
\Big\} \nnb \\
%10
\ar 8 m_{\Lambda_b}^2
 \hat{s}^2 \Big(1-\sqrt{\hat{r}_\Lambda}\Big)
\Big[ 2 (A_2-B_2) C_{TE}^\ast - (D_2-E_2) C_{T}^\ast \Big] 
f_T^{V\ast} \nnb \\
%11
\ek 16 m_{\Lambda_b}^2 \hat{m}_\ell   
 \hat{s} \Big(1+2 \sqrt{\hat{r}_\Lambda}+\hat{r}_\Lambda
-\hat{s}\Big) F_1^\ast C_{T} f_T^S
\Bigg\}~, \\ \nnb \\
\label{e6624}
P_{NT} \es \frac{16 m_{\Lambda_b}^4 v}{3 \Delta}
\mbox{\rm Im} \Bigg\{
%1
4 \lambda  
(A_1 D_1^\ast +B_1 E_1^\ast) \nnb \\
%2
\ar 32 m_{\Lambda_b} \hat{m}_\ell \lambda 
\Big[ (D_1 + E_1) C_{T}^\ast f_T^{V\ast} - (D_2 + E_2) C_{T}^\ast f_T^\ast 
+ 2 (D_2 - E_2) C_{TE}^\ast f_T^\ast\Big] \nnb \\
%3
\ar 32 m_{\Lambda_b}^2 \hat{m}_\ell \lambda   
\Big( 1+\sqrt{\hat{r}_\Lambda} \Big)
\Big[(A_1+B_1) C_{TE}^\ast f_T^{S\ast} +
(D_1+E_1) C_{T}^\ast f_T^{S\ast} \Big] \nnb \\
%4
\ar 6 \hat{m}_\ell \Big(1+\sqrt{\hat{r}_\Lambda}\Big)
\Big(1-2 \sqrt{\hat{r}_\Lambda} + \hat{r}_\Lambda - \hat{s}\Big)
(D_1-E_1) H_1^\ast \nnb \\
%5
\ek 6 \hat{m}_\ell \Big(1-\sqrt{\hat{r}_\Lambda}\Big)
\Big(1+2 \sqrt{\hat{r}_\Lambda} + \hat{r}_\Lambda - \hat{s}\Big)
(D_1+E_1) F_1^\ast \nnb \\
%6
\ek 32 m_{\Lambda_b}^3 \hat{m}_\ell \lambda \hat{s}
\Big[ (A_2+B_2) C_{TE}^\ast + (D_2+E_2) C_{T}^\ast \Big] f_T^{S\ast} \nnb \\
%7
\ar 4 m_{\Lambda_b}^2 \lambda \hat{s}
(A_2^\ast D_2 + B_2^\ast E_2) \nnb \\
%8
\ar 256 m_{\Lambda_b}^2 \lambda \hat{s}
\Big\{ \mbox{\rm Re}\Big[f_T f_T^{S\ast}\Big] - 
m_{\Lambda_b} \Big(1+\sqrt{\hat{r}_\Lambda}\Big)   
\mbox{\rm Re}\Big[f_T^S f_T^{V\ast}\Big] \Big\} C_{T} C_{TE}^\ast \nnb \\
%9
\ar 3 \lambda \hat{s}  
\Big(1-2 \sqrt{\hat{r}_\Lambda} + \hat{r}_\Lambda - \hat{s}\Big)
\Big[ H_1 H_2^\ast - 2 m_{\Lambda_b} \hat{m}_\ell 
(D_3-E_3) H_1^\ast \Big] \nnb \\
%10
\ar 96 m_{\Lambda_b} \hat{s} \Big(1+\sqrt{\hat{r}_\Lambda}\Big)
\Big(1-2 \sqrt{\hat{r}_\Lambda} + \hat{r}_\Lambda - \hat{s}\Big)
\Big\{ 8 \mbox{\rm Re}\Big[f_T f_T^{V\ast}\Big] C_{T} C_{TE}^\ast \nnb \\
\ek m_{\Lambda_b} \hat{m}_\ell (A_2+B_2) 
C_{TE}^\ast f_T^{V\ast} \Big\} \nnb \\
%11
\ek \hat{s}  
\Big(1+2 \sqrt{\hat{r}_\Lambda} + \hat{r}_\Lambda - \hat{s}\Big)
\Big\{ 128 \lambda m_{\Lambda_b}^4 \vel f_T^S \ver^2 
C_{T} C_{TE}^\ast \nnb \\
\ar 3 \Big[F_2 + 2 m_{\Lambda_b} \hat{m}_\ell (D_3+E_3)\Big] 
F_1^\ast \Big\} \nnb \\
%12
\ek 32 m_{\Lambda_b}          
\Big[ (1-\hat{r}_\Lambda)^2 + \hat{s} \Big(1-6 \sqrt{\hat{r}_\Lambda} + 
\hat{r}_\Lambda\Big) -2 \hat{s}^2 \Big] \nnb \\ 
\cp \Big[ 4 m_{\Lambda_b} \hat{s} \vel f_T^V \ver^2 
C_{T} C_{TE}^\ast - \hat{m}_\ell (A_1+B_1) 
C_{TE}^\ast f_T^{V\ast} \Big] \nnb \\
%13
\ar 1536 \sqrt{\hat{r}_\Lambda} \hat{s} 
\vel f_T \ver^2 C_{T} C_{TE}^\ast \nnb \\
%14
\ek 48 \hat{m}_\ell
\Big\{ (1-\hat{r}_\Lambda) \Big[ C_{T}-2 C_{TE} + \sqrt{\hat{r}_\Lambda}
(C_{T}+2 C_{TE}) \Big] \nnb \\
\ek \hat{s} \Big[ C_{T}-2 C_{TE} - \sqrt{\hat{r}_\Lambda}
(C_{T}+2 C_{TE}) \Big] \Big\} B_1^\ast f_T \nnb \\
%15
\ar 16 m_{\Lambda_b} \hat{m}_\ell
\Big\{ 2(1-\hat{r}_\Lambda)^2 (C_{T}-2 C_{TE}) \nnb \\
\ek \hat{s} \Big[ \Big(1-6 \sqrt{\hat{r}_\Lambda} +\hat{r}_\Lambda\Big) C_{T}
- 2 \Big(1+6 \sqrt{\hat{r}_\Lambda} +\hat{r}_\Lambda\Big) C_{TE} \Big]
- \hat{s}^2 (C_{T}-2 C_{TE}) \Big\} A_2^\ast f_T \nnb \\
%16
\ek 16m_{\Lambda_b} \hat{m}_\ell
\Big\{ 2(1-\hat{r}_\Lambda)^2 (C_{T}+2 C_{TE}) \nnb \\
\ek \hat{s} \Big[ \Big(1-6 \sqrt{\hat{r}_\Lambda} +\hat{r}_\Lambda\Big) C_{T} +
2 \Big(1+6 \sqrt{\hat{r}_\Lambda} +\hat{r}_\Lambda\Big) C_{TE} \Big]
- \hat{s}^2 (C_{T}+2 C_{TE}) \Big\} B_2^\ast f_T \nnb \\
%17
\ar 48 \hat{m}_\ell
\Big\{ (1-\hat{r}_\Lambda) \Big[ C_{T} + 2 C_{TE} +   
\sqrt{\hat{r}_\Lambda} (C_{T} - 2 C_{TE}) \Big] \nnb \\
\ek \hat{s} \Big[ C_{T} + 2 C_{TE} -              
\sqrt{\hat{r}_\Lambda} (C_{T} - 2 C_{TE}) \Big] \Big\} A_1^\ast f_T
\Bigg\}~, \\ \nnb \\
\label{e6625}
P_{TN} \es - \frac{16 m_{\Lambda_b}^4 v}{3 \Delta}
\mbox{\rm Im} \Bigg\{
%1
4 \lambda 
(A_1 D_1^\ast +B_1 E_1^\ast) \nnb \\
%2
\ar 32 m_{\Lambda_b} \hat{m}_\ell \lambda
\Big[ (D_1 + E_1) C_{T}^\ast f_T^{V\ast} - (D_2 + E_2) C_{T}^\ast f_T^\ast
+ 2 (D_2 - E_2) C_{TE}^\ast f_T^\ast\Big] \nnb \\
%3
\ek 32 m_{\Lambda_b}^2 \hat{m}_\ell \lambda   
\Big( 1+\sqrt{\hat{r}_\Lambda} \Big)
\Big[(A_1+B_1) C_{TE}^\ast f_T^{S\ast} -
(D_1+E_1) C_{T}^\ast f_T^{S\ast} \Big] \nnb \\
%4
\ek 6 \hat{m}_\ell \Big(1+\sqrt{\hat{r}_\Lambda}\Big)
\Big(1-2 \sqrt{\hat{r}_\Lambda} + \hat{r}_\Lambda - \hat{s}\Big)
(D_1-E_1) H_1^\ast \nnb \\
%5
\ar 6 \hat{m}_\ell \Big(1-\sqrt{\hat{r}_\Lambda}\Big)
\Big(1+2 \sqrt{\hat{r}_\Lambda} + \hat{r}_\Lambda - \hat{s}\Big)
(D_1+E_1) F_1^\ast \nnb \\
%6
\ar 32 m_{\Lambda_b}^3 \hat{m}_\ell \lambda \hat{s}
\Big[ (A_2+B_2) C_{TE}^\ast - (D_2+E_2) C_{T}^\ast \Big] f_T^{S\ast} \nnb \\
%7
\ar 4 m_{\Lambda_b}^2 \lambda \hat{s}
(A_2^\ast D_2 + B_2^\ast E_2) \nnb \\
%8
\ek 256 m_{\Lambda_b}^2 \lambda \hat{s}
\Big\{ \mbox{\rm Re}\Big[f_T f_T^{S\ast}\Big] - 
m_{\Lambda_b} \Big(1+\sqrt{\hat{r}_\Lambda}\Big)   
\mbox{\rm Re}\Big[f_T^S f_T^{V\ast}\Big] \Big\} C_{T} C_{TE}^\ast \nnb \\
%9
\ek 3 \lambda \hat{s}  
\Big(1-2 \sqrt{\hat{r}_\Lambda} + \hat{r}_\Lambda - \hat{s}\Big)
\Big[ H_1 H_2^\ast - 2 m_{\Lambda_b} \hat{m}_\ell 
(D_3-E_3) H_1^\ast \Big] \nnb \\
%10
\ek 96 m_{\Lambda_b} \hat{s} \Big(1+\sqrt{\hat{r}_\Lambda}\Big)
\Big(1-2 \sqrt{\hat{r}_\Lambda} + \hat{r}_\Lambda - \hat{s}\Big)
\Big\{ 8 \mbox{\rm Re}\Big[f_T f_T^{V\ast}\Big] C_{T} C_{TE}^\ast \nnb \\
\ek m_{\Lambda_b} \hat{m}_\ell (A_2+B_2) 
C_{TE}^\ast f_T^{V\ast} \Big\} \nnb \\
%11
\ar \hat{s}
\Big(1+2 \sqrt{\hat{r}_\Lambda} + \hat{r}_\Lambda - \hat{s}\Big)
\Big\{ 128 \lambda m_{\Lambda_b}^4 \vel f_T^S \ver^2 
C_{T} C_{TE}^\ast \nnb \\ 
\ar 3 \Big[F_2 + 2 m_{\Lambda_b} \hat{m}_\ell (D_3+E_3)\Big]
F_1^\ast \Big\} \nnb \\
%12
\ar 32 m_{\Lambda_b}          
\Big[ (1-\hat{r}_\Lambda)^2 + \hat{s} \Big(1-6 \sqrt{\hat{r}_\Lambda} + 
\hat{r}_\Lambda\Big) -2 \hat{s}^2 \Big] \nnb \\ 
\cp \Big[ 4 m_{\Lambda_b} \hat{s} \vel f_T^V \ver^2 
C_{T} C_{TE}^\ast - \hat{m}_\ell (A_1+B_1) 
C_{TE}^\ast f_T^{V\ast} \Big] \nnb \\
%13
\ek 1536 \sqrt{\hat{r}_\Lambda} \hat{s}
\vel f_T \ver^2 C_{T} C_{TE}^\ast \nnb \\
%14
\ar 48 \hat{m}_\ell
\Big\{ (1-\hat{r}_\Lambda) \Big[ C_{T}-2 C_{TE} + \sqrt{\hat{r}_\Lambda}
(C_{T}+2 C_{TE}) \Big] \nnb \\ 
\ek \hat{s} \Big[ C_{T}-2 C_{TE} - \sqrt{\hat{r}_\Lambda}
(C_{T}+2 C_{TE}) \Big] \Big\} B_1^\ast f_T \nnb \\
%15
\ek 16 m_{\Lambda_b} \hat{m}_\ell
\Big\{ 2(1-\hat{r}_\Lambda)^2 (C_{T}-2 C_{TE}) \nnb \\
\ek \hat{s} \Big[ \Big(1-6 \sqrt{\hat{r}_\Lambda} +\hat{r}_\Lambda\Big) C_{T}
- 2 \Big(1+6 \sqrt{\hat{r}_\Lambda} +\hat{r}_\Lambda\Big) C_{TE} \Big]
- \hat{s}^2 (C_{T}-2 C_{TE}) \Big\} A_2^\ast f_T \nnb \\
%16
\ar 16 m_{\Lambda_b} \hat{m}_\ell
\Big\{ 2(1-\hat{r}_\Lambda)^2 (C_{T}+2 C_{TE}) \nnb \\
\ek \hat{s} \Big[ \Big(1-6 \sqrt{\hat{r}_\Lambda} +\hat{r}_\Lambda\Big) C_{T} +
2 \Big(1+6 \sqrt{\hat{r}_\Lambda} +\hat{r}_\Lambda\Big) C_{TE} \Big]
- \hat{s}^2 (C_{T}+2 C_{TE}) \Big\} B_2^\ast f_T \nnb \\
%17
\ek 48 \hat{m}_\ell 
\Big\{ (1-\hat{r}_\Lambda) \Big[ C_{T} + 2 C_{TE} +   
\sqrt{\hat{r}_\Lambda} (C_{T} - 2 C_{TE}) \Big] \nnb \\
\ek \hat{s} \Big[ C_{T} + 2 C_{TE} -              
\sqrt{\hat{r}_\Lambda} (C_{T} - 2 C_{TE}) \Big] \Big\} A_1^\ast f_T
\Bigg\}~, \\ \nnb \\
\label{e6626}
P_{NN} \es \frac{8 m_{\Lambda_b}^4}{3 \hat{s} \Delta}
\mbox{\rm Re} \Bigg\{
%1
96 m_{\Lambda_b}^2 \hat{m}_\ell \lambda \hat{s}    
\Big( 1+ \sqrt{\hat{r}_\Lambda} \Big)
(A_1+B_1) C_{T}^\ast f_T^{S\ast} \nnb \\
%2
\ar 96 \hat{m}_\ell^2 \sqrt{\hat{r}_\Lambda} \hat{s}
( A_1 B_1^\ast + D_1 E_1^\ast ) \nnb \\
%3
\ek 48 m_{\Lambda_b} \hat{m}_\ell^2 \sqrt{\hat{r}_\Lambda} \hat{s}  
(1-\hat{r}_\Lambda +\hat{s}) (A_1 A_2^\ast + B_1 B_2^\ast) \nnb \\
%4
\ar 96 m_{\Lambda_b} \hat{m}_\ell \hat{s} \Big( 1 + \sqrt{\hat{r}_\Lambda}
\Big)^2
\Big( 1 - 2 \sqrt{\hat{r}_\Lambda} + \hat{r}_\Lambda -\hat{s} \Big)
(A_1 + B_1) C_{T}^\ast f_T^{V\ast} \nnb \\
%5
\ar 12 \hat{m}_\ell \hat{s} \Big( 1 - \sqrt{\hat{r}_\Lambda} \Big)
\Big( 1 + 2 \sqrt{\hat{r}_\Lambda} + \hat{r}_\Lambda -\hat{s} \Big)
(D_1 + E_1) F_2^\ast \nnb \\
%6
\ek 96 m_{\Lambda_b}^3 \hat{m}_\ell \lambda \hat{s}^2
(A_2 + B_2) C_{T}^\ast f_T^{S\ast} \nnb \\
%7
\ek 12 \hat{m}_\ell \hat{s} \Big( 1 + \sqrt{\hat{r}_\Lambda} \Big)  
\Big( 1 - 2 \sqrt{\hat{r}_\Lambda} + \hat{r}_\Lambda -\hat{s} \Big)
\Big[ (D_1-E_1) H_2^\ast + 8 m_{\Lambda_b}^2 \hat{s} (A_2 + B_2)
C_{T}^\ast f_T^{V\ast} \Big] \nnb \\
%8
\ar 3 \hat{s}^2 
\Big( 1 + 2 \sqrt{\hat{r}_\Lambda} + \hat{r}_\Lambda -\hat{s} \Big)
\Big[ \vel F_2 \ver^2 + 4 m_{\Lambda_b} \hat{m}_\ell
(D_3 + E_3 ) F_2^\ast \Big] \nnb \\
%9
\ar 24 m_{\Lambda_b} \hat{m}_\ell^2 \hat{s} \Big[ 
m_{\Lambda_b} \hat{s} (1+\hat{r}_\Lambda-\hat{s})
\Big(\vel D_3 \ver^2 + \vel E_3 \ver^2 \Big) +
2 \sqrt{\hat{r}_\Lambda} (1-\hat{r}_\Lambda+\hat{s})
(D_1 D_3^\ast + E_1 E_3^\ast)\Big] \nnb \\        
%10
\ar 48 m_{\Lambda_b} \hat{m}_\ell^2 \hat{s} (1-\hat{r}_\Lambda-\hat{s})
(A_1 B_2^\ast + A_2 B_1 ^\ast + D_1 E_3^\ast + D_3 E_1^\ast) \nnb \\
%11
\ek 4 [ \lambda \hat{s} + 
2 \hat{m}_\ell^2 (1 + \hat{r}_\Lambda^2 - 2 \hat{r}_\Lambda + 
\hat{r}_\Lambda \hat{s} + \hat{s} - 2 \hat{s}^2) ]
\Big( \vel A_1 \ver^2 + \vel B_1 \ver^2 - \vel D_1 \ver^2 - 
\vel E_1 \ver^2 \Big) \nnb \\
%12
\ar 3 \hat{s}^2 \Big( 1-2 \sqrt{\hat{r}_\Lambda} +
\hat{r}_\Lambda -\hat{s}\Big) \Big[ 4 m_{\Lambda_b} \hat{m}_\ell
(D_3-E_3) H_2^\ast + \vel H_2 \ver^2 \Big] \nnb \\
%13
\ar 96 m_{\Lambda_b}^2 \hat{m}_\ell^2 \sqrt{\hat{r}_\Lambda} \hat{s}^2 
(A_2 B_2^\ast + D_3 E_3^\ast) \nnb \\
%14
\ar 64 m_{\Lambda_b}^2 \lambda \hat{s}^2
\Big[ (3 - 2 v^2) \vel C_{T} \ver^2 -   
4 v^2 \vel C_{TE} \ver^2 \Big]
\Big\{2 f_T f_T^{S\ast} - m_{\Lambda_b} \Big[ 
2 \Big(1+\sqrt{\hat{r}_\Lambda}\Big) f_T^S f_T^{V\ast} \nnb \\
\ar 
m_{\Lambda_b}\Big(1+2 \sqrt{\hat{r}_\Lambda}+ \hat{r}_\Lambda - \hat{s}\Big)
\vel f_T^S\ver^2 \Big] \Big\} \nnb \\
%15
\ek 4 m_{\Lambda_b}^2 \lambda \hat{s}^2 v^2 
\Big( \vel D_2 \ver^2 + \vel E_2 \ver^2 \Big) \nnb \\
%16
\ar 4 m_{\Lambda_b}^2 \hat{s} \{ \lambda \hat{s} -
2 \hat{m}_\ell^2 [2 (1+ \hat{r}_\Lambda^2) - \hat{s} (1+\hat{s})
- \hat{r}_\Lambda (4+\hat{s})]\}
\Big( \vel A_2 \ver^2 + \vel B_2 \ver^2 \Big) \nnb \\
%17
\ar 96 m_{\Lambda_b} \hat{m}_\ell \hat{s}^2 
\Big[ \Big(1-2 \sqrt{\hat{r}_\Lambda} + \hat{r}_\Lambda - \hat{s}\Big)
B_2^\ast f_T C_{T} +   
2 \Big(1+2 \sqrt{\hat{r}_\Lambda} + \hat{r}_\Lambda - \hat{s}\Big)
B_2^\ast f_T C_{TE} \Big] \nnb \\
%18
\ek 96 \hat{m}_\ell \hat{s}        
\Big[ \Big( 1+\sqrt{\hat{r}_\Lambda} \Big)
\Big(1-2 \sqrt{\hat{r}_\Lambda} + \hat{r}_\Lambda - \hat{s}\Big)
A_1^\ast f_T C_{T} \nnb \\
\ar 2 \Big( 1-\sqrt{\hat{r}_\Lambda} \Big)
\Big(1+2 \sqrt{\hat{r}_\Lambda} + \hat{r}_\Lambda - \hat{s}\Big)
A_1^\ast f_T C_{TE} \Big] \nnb \\
%19
\ek 96 \hat{m}_\ell \hat{s}
\Big[ \Big( 1+\sqrt{\hat{r}_\Lambda} \Big)
\Big(1-2 \sqrt{\hat{r}_\Lambda} + \hat{r}_\Lambda - \hat{s}\Big)
B_1^\ast f_T C_{T} \nnb \\
\ek 2 \Big( 1-\sqrt{\hat{r}_\Lambda} \Big)
\Big(1+2 \sqrt{\hat{r}_\Lambda} + \hat{r}_\Lambda - \hat{s}\Big)
B_1^\ast f_T C_{TE} \Big] \nnb \\
%20
\ek 768 \hat{s}
\Big\{ \Big[ \hat{m}_\ell^2 \Big(1 + 2 \sqrt{\hat{r}_\Lambda} +
\hat{r}_\Lambda   
-\hat{s} \Big) - \sqrt{\hat{r}_\Lambda} \hat{s}\Big] 
\vel C_{T} \ver^2 \nnb \\
\ar 4 \Big[ \hat{m}_\ell^2 \Big(1 - 2 \sqrt{\hat{r}_\Lambda} + \hat{r}_\Lambda   
-\hat{s} \Big) + \sqrt{\hat{r}_\Lambda} \hat{s}\Big] \vel C_{TE} \ver^2
\Big\} \vel f_T \ver^2 \nnb \\
%21
\ar 96 m_{\Lambda_b} \hat{m}_\ell \hat{s}^2
\Big[ \Big(1-2 \sqrt{\hat{r}_\Lambda} + \hat{r}_\Lambda - \hat{s}\Big)
C_{T}^\ast -                                       
2 \Big(1+2 \sqrt{\hat{r}_\Lambda} + \hat{r}_\Lambda - \hat{s}\Big)
C_{TE}^\ast \Big] A_2 f_T^\ast \nnb \\
%22
\ek 3 \hat{s}^2 v^2
\Big[ \Big(1+2 \sqrt{\hat{r}_\Lambda} + \hat{r}_\Lambda - \hat{s}\Big)
\vel F_1 \ver^2 + 
\Big(1-2 \sqrt{\hat{r}_\Lambda} + \hat{r}_\Lambda - \hat{s}\Big)
\vel H_1 \ver^2 \Big] \nnb \\
%23
\ar 384 m_{\Lambda_b} \hat{s}^2 \Big( 1+\sqrt{\hat{r}_\Lambda} \Big)
\Big(1-2 \sqrt{\hat{r}_\Lambda} + \hat{r}_\Lambda - \hat{s}\Big)
\Big( \vel C_{T} \ver^2 - 4 v^2 \vel C_{TE} \ver^2 \Big)
f_T f_T^{V\ast} \nnb \\
%24
\ek 64 m_{\Lambda_b}^2 \hat{s} \Big(1-2 \sqrt{\hat{r}_\Lambda} + 
\hat{r}_\Lambda - \hat{s}\Big)
\Big\{ \Big[8 \hat{m}_\ell^2 \Big(1+2 \sqrt{\hat{r}_\Lambda} +
\hat{r}_\Lambda -  \hat{s}\Big) \nnb \\
\ar \hat{s} \Big(1+2 \sqrt{\hat{r}_\Lambda} + \hat{r}_\Lambda +  
2 \hat{s}\Big)\Big] \vel C_{T} \ver^2 -
4 \hat{s} v^2 \Big(1+2 \sqrt{\hat{r}_\Lambda} + \hat{r}_\Lambda +  
2 \hat{s}\Big) \vel C_{TE} \ver^2 \Big\} \vel f_T^V \ver^2
\Bigg\}~, \\ \nnb \\
\label{e6627}
P_{TT} \es \frac{8 m_{\Lambda_b}^4}{3 \hat{s} \Delta}
\mbox{\rm Re} \Bigg\{
%1
32 m_{\Lambda_b}^2 \hat{m}_\ell \lambda \hat{s}    
\Big( 1+ \sqrt{\hat{r}_\Lambda} \Big)
(A_1+B_1) C_{T}^\ast f_T^{S\ast}\nnb \\
%2
\ek 96 \hat{m}_\ell^2 \sqrt{\hat{r}_\Lambda} \hat{s}
( A_1 B_1^\ast + D_1 E_1^\ast ) \nnb \\
%3
\ek 48 m_{\Lambda_b} \hat{m}_\ell^2 \sqrt{\hat{r}_\Lambda} \hat{s}  
(1-\hat{r}_\Lambda +\hat{s}) (D_1 D_3^\ast + E_1 E_3^\ast) \nnb \\
%4
\ek 12 \hat{m}_\ell \hat{s} \Big( 1 - \sqrt{\hat{r}_\Lambda} \Big)  
\Big( 1 + 2 \sqrt{\hat{r}_\Lambda} + \hat{r}_\Lambda -\hat{s} \Big)
(D_1 + E_1) F_2^\ast \nnb \\
%5
\ek 32 m_{\Lambda_b}^3 \hat{m}_\ell \lambda \hat{s}^2
(A_2 + B_2) C_{T}^\ast f_T^{S\ast} \nnb \\
%6
\ek 96 m_{\Lambda_b}^2 \hat{m}_\ell^2 \sqrt{\hat{r}_\Lambda} \hat{s}^2
( A_2 B_2^\ast + D_3 E_3^\ast ) \nnb \\
%7
\ar 12 \hat{m}_\ell \hat{s} \Big( 1 + \sqrt{\hat{r}_\Lambda} \Big)  
\Big( 1 - 2 \sqrt{\hat{r}_\Lambda} + \hat{r}_\Lambda -\hat{s} \Big)
\Big[ (D_1-E_1) H_2^\ast + 8 m_{\Lambda_b}^2 \hat{s} (A_2 + B_2)
C_{T}^\ast f_T^{V\ast} \Big] \nnb \\
%8
\ek 3 \hat{s}^2
\Big( 1 + 2 \sqrt{\hat{r}_\Lambda} + \hat{r}_\Lambda -\hat{s} \Big)
\Big[ \vel F_2 \ver^2 + 4 m_{\Lambda_b} \hat{m}_\ell               
(D_3 + E_3 ) F_2^\ast \Big] \nnb \\
%9
\ek 24 m_{\Lambda_b} \hat{m}_\ell^2 \hat{s} \Big[ 
m_{\Lambda_b} \hat{s} (1+\hat{r}_\Lambda-\hat{s})
\Big(\vel D_3 \ver^2 + \vel E_3 \ver^2 \Big) -
2 \sqrt{\hat{r}_\Lambda} (1-\hat{r}_\Lambda+\hat{s})
(A_1 A_2^\ast + B_1 B_2^\ast)\Big] \nnb \\
%10
\ek 48 m_{\Lambda_b} \hat{m}_\ell^2 \hat{s} (1-\hat{r}_\Lambda-\hat{s})
(A_1 B_2^\ast + A_2 B_1 ^\ast + D_1 E_3^\ast + D_3 E_1^\ast) \nnb \\
%11
\ek 4 [ \lambda \hat{s} - 
2 \hat{m}_\ell^2 (1 + \hat{r}_\Lambda^2 - 2 \hat{r}_\Lambda + 
\hat{r}_\Lambda \hat{s} + \hat{s} - 2 \hat{s}^2) ]
\Big( \vel A_1 \ver^2 + \vel B_1 \ver^2 \Big) \nnb \\
%12
\ar 4 m_{\Lambda_b}^2 \hat{s} \{ \lambda \hat{s}^2 +
\hat{m}_\ell^2 [4 (1- \hat{r}_\Lambda)^2 - 2 \hat{s} (1+\hat{r}_\Lambda)         
- 2 \hat{s}^2) ]\}
\Big( \vel A_2 \ver^2 + \vel B_2 \ver^2 \Big) \nnb \\
%13
\ar 4 \{ \lambda \hat{s} -
2 \hat{m}_\ell^2 [5 (1- \hat{r}_\Lambda)^2 - 7 \hat{s} (1+\hat{r}_\Lambda)         
+ 2 \hat{s}^2) ]\}                                                               
\Big( \vel D_1 \ver^2 + \vel E_1 \ver^2 \Big) \nnb \\
%14
\ar 32 m_{\Lambda_b} \hat{m}_\ell \hat{s}                                   
\Big[(1- \hat{r}_\Lambda)^2 - \hat{s} (5 - 6 \sqrt{\hat{r}_\Lambda} + 
5 \hat{r}_\Lambda\Big) + 4 \hat{s}^2 \Big]
\Big[ (A_1 + B_1 ) C_{T} f_T^{V\ast}\Big] \nnb \\
%15
\ek 3 \hat{s}^2 \Big( 1-2 \sqrt{\hat{r}_\Lambda} +
\hat{r}_\Lambda -\hat{s}\Big) \Big[ 4 m_{\Lambda_b} \hat{m}_\ell
(D_3-E_3) H_2^\ast + \vel H_2 \ver^2 \Big] \nnb \\
%16
\ek 64 m_{\Lambda_b}^2 \lambda \hat{s}^2
\Big[ (2 v^2-1) \vel C_{T} \ver^2 -
4 v^2 \vel C_{TE} \ver^2 \Big]
\Big\{2 f_T f_T^{S\ast} - m_{\Lambda_b} \Big[
2 \Big(1+\sqrt{\hat{r}_\Lambda}\Big) f_T^S f_T^{V\ast} \nnb \\
\ar m_{\Lambda_b}\Big(1+2 \sqrt{\hat{r}_\Lambda}+ \hat{r}_\Lambda - \hat{s}\Big)
\vel f_T^S\ver^2 \Big] \Big\} \nnb \\
%17
\ek 32 m_{\Lambda_b} \hat{m}_\ell \hat{s} \Big\{
[4 (1-\hat{r}_\Lambda)^2 - 5 \hat{s} (1+\hat{r}_\Lambda)+\hat{s}^2] 
A_2^\ast (C_{T}-2 C_{TE}) f_T -
6 \sqrt{\hat{r}_\Lambda} \hat{s} A_2^\ast (C_{T}+2 C_{TE}) 
f_T \Big\} \nnb \\
%18
\ek 32 m_{\Lambda_b} \hat{m}_\ell \hat{s} \Big\{
[4 (1-\hat{r}_\Lambda)^2 - 5 \hat{s} (1+\hat{r}_\Lambda)+\hat{s}^2]
B_2^\ast (C_{T}+2 C_{TE}) f_T -
6 \sqrt{\hat{r}_\Lambda} \hat{s} B_2^\ast (C_{T}-2 C_{TE}) 
f_T \Big\} \nnb \\
%19
\ar 256
\Big\{ - 3 \sqrt{\hat{r}_\Lambda} \hat{s}^2 +            
\hat{m}_\ell^2 \Big[ 4 (1-\hat{r}_\Lambda)^2 -
\hat{s} \Big(5 - 6 \sqrt{\hat{r}_\Lambda} + 5 \hat{r}_\Lambda \Big) +
\hat{s}^2 \Big] \Big\} \vel C_{T} \ver^2 \vel f_T \ver^2 \nnb \\
%20
\ar 1024
\Big\{3 \sqrt{\hat{r}_\Lambda} \hat{s}^2 +
\hat{m}_\ell^2 \Big[ 4 (1-\hat{r}_\Lambda)^2 -
\hat{s} \Big(5 + 6 \sqrt{\hat{r}_\Lambda} + 5 \hat{r}_\Lambda \Big) +
\hat{s}^2 \Big] \Big\} \vel C_{TE} \ver^2 \vel f_T \ver^2 \nnb \\
%21
\ar 96 \hat{m}_\ell \hat{s}
\Big[ \Big( 1+\sqrt{\hat{r}_\Lambda} \Big)
\Big(1-2 \sqrt{\hat{r}_\Lambda} + \hat{r}_\Lambda - \hat{s}\Big)
C_{T}^\ast \nnb \\
\ar 2 \Big( 1-\sqrt{\hat{r}_\Lambda} \Big)
\Big(1+2 \sqrt{\hat{r}_\Lambda} + \hat{r}_\Lambda - \hat{s}\Big)
C_{TE}^\ast \Big] A_1 f_T^\ast \nnb \\
%22
\ar 96 \hat{m}_\ell \hat{s}
\Big[ \Big( 1+\sqrt{\hat{r}_\Lambda} \Big)
\Big(1-2 \sqrt{\hat{r}_\Lambda} + \hat{r}_\Lambda - \hat{s}\Big)
C_{T}^\ast \nnb \\
\ek 2 \Big( 1-\sqrt{\hat{r}_\Lambda} \Big)
\Big(1+2 \sqrt{\hat{r}_\Lambda} + \hat{r}_\Lambda - \hat{s}\Big)
C_{TE}^\ast \Big] B_1 f_T^\ast \nnb \\
%23
\ek 4 m_{\Lambda_b}^2 \lambda \hat{s}^2 v^2 
\Big( \vel D_2 \ver^2 + \vel E_2 \ver^2 \Big) \nnb \\
%24
\ar 3 \hat{s}^2 v^2 \Big[
\Big(1+2 \sqrt{\hat{r}_\Lambda} + \hat{r}_\Lambda - \hat{s}\Big) 
\vel F_1 \ver^2 +
\Big(1-2 \sqrt{\hat{r}_\Lambda} + \hat{r}_\Lambda - \hat{s}\Big) 
\vel H_1 \ver^2 \Big] \nnb \\
%25
\ek 384 m_{\Lambda_b} \hat{s}^2 \Big( 1+\sqrt{\hat{r}_\Lambda} \Big)
\Big(1-2 \sqrt{\hat{r}_\Lambda} + \hat{r}_\Lambda - \hat{s}\Big)
\Big( \vel C_{T} \ver^2 - 4 v^2 \vel C_{TE} \ver^2 \Big)
f_T f_T^{V\ast} \nnb \\
%26
\ek 64 m_{\Lambda_b}^2 \hat{s} \Big(1-2 \sqrt{\hat{r}_\Lambda} +
\hat{r}_\Lambda - \hat{s}\Big)
\Big\{ \Big[8 \hat{m}_\ell^2 \Big(1+2 \sqrt{\hat{r}_\Lambda} +
\hat{r}_\Lambda - \hat{s}\Big) \nnb \\
\ek 
\hat{s} \Big(1+2 \sqrt{\hat{r}_\Lambda} + \hat{r}_\Lambda +
2 \hat{s}\Big)\Big] \vel C_{T} \ver^2 +
4 \hat{s} v^2 \Big(1+2 \sqrt{\hat{r}_\Lambda} + \hat{r}_\Lambda +
2 \hat{s}\Big) \vel C_{TE} \ver^2 \Big\} \vel f_T^V \ver^2
\Bigg\}~.
\eea

\section{Numerical analysis}

We start this section by presenting the numerical results for all 
possible double--lepton
polarization asymmetries. The values of the input parameters we use in out
calculations are: $\vel V_{tb} V_{ts}^\ast \ver = 0.0385$, 
$m_\tau = 1.77~GeV$, $m_\mu = 0.106~GeV$.
$m_b = 4.8~GeV$. For the Wilson coefficients we use their SM values which
are given as:
$C_7^{SM} = -0.313$, $C_9^{SM} = 4.344$ and $C_{10}^{SM} = -4.669$.
The magnitude of $C_7^{eff}$ is quite well constrained from $b \rar s
\gamma$ decay, and hence well established. Moreover, we will fix the values 
of the Wilson coefficients,
i.e., $C_{BR}$ and $C_{SL}$ are both related to $C_7^{eff}$ as follows:
$C_{BR}=-2 m_b C_7^{eff}$ and $C_{SL}=-2 m_s C_7^{eff}$.  
As far as the Wilson coefficient
$C_9^{SM}$ is considered, we take into account only the short distance 
contributions, and we neglect the 
long distance contributions, coming from the production of
$\bar{c}c$ pair at intermediate states. It is well known that the form
factors are the main and the most important 
input parameters necessary in
the numerical calculations. The calculation of the form factors of $\Lambda_b
\rar \Lambda$ transition does not exist at present.
But, we can use the results from QCD sum rules
in corporation with HQET \cite{R6622,R6623}. We noted earlier that,
HQET allows us to establish relations among the form factors and reduces
the number of independent form factors into two. 
In \cite{R6622,R6623}, the $q^2$ dependence of these form factors
are given as follows
\bea
F(\hat{s}) = \frac{F(0)}{\ds 1-a_F \hat{s} + b_F \hat{s}^2}~. \nnb
\eea
The values of the parameters $F(0),~a_F$ and $b_F$ are given in table 1.
\begin{table}[h]    
\renewcommand{\arraystretch}{1.5}
\addtolength{\arraycolsep}{3pt}
$$
\begin{array}{|l|ccc|}  
\hline
& F(0) & a_F & b_F \\ \hline
F_1 &
\phantom{-}0.462 & -0.0182 & -0.000176 \\
F_2 &
-0.077 & -0.0685 &\phantom{-}0.00146 \\ \hline
\end{array}
$$
\caption{Form factors for $\Lambda_b \rar \Lambda \ell^+ \ell^-$
decay in a three parameter fit.}
\renewcommand{\arraystretch}{1}
\addtolength{\arraycolsep}{-3pt}
\end{table}  

In further numerical analysis, the values of the new Wilson coefficients
which describe new physics beyond the SM, are needed. In numerical
calculations we will vary all new Wilson coefficients in the range
$- \vel C_{10}^{SM} \ver \le C_X \le \vel C_{10}^{SM} \ver$. The
experimental results on the branching ratio of the 
$B \rar K^\ast \ell^+ \ell^-$ decay \cite{R6613,R6614} and the bound on the
branching ratio of $B \rar \mu^+ \mu^-$ \cite{R6624} suggest that 
this is the right order of
magnitude for the vector and scalar interaction coefficients. Here, we
emphasize that the existing experimental results on the 
$B \rar K^\ast \ell^+ \ell^-$ and $B \rar K \ell^+ \ell^-$ decays put
stronger restrictions on some of the new Wilson coefficients. For example,
$-2 \le C_{LL} \le 0$, $0 \le C_{RL} \le 2.3$, $-1.5 \le C_{T} \le 1.5$ and
$-3.3 \le C_{TE} \le 2.6$, and all of the remaining Wilson coefficients vary
in the region $- \vel C_{10}^{SM} \ver \le C_X \le \vel C_{10}^{SM} \ver$.

It follows from the explicit expressions of the double--lepton polarization
asymmetries that they depend on $q^2$ and the new Wilson coefficients. For
this reason there may appear difficulties in studying the dependencies of
the physical properties on both parameters at the same time. Hence, it is
necessary to eliminate the dependence of $P_{ij}$ on one of these
parameters. Here in the present work, we eliminate $q^2$ dependence of
$P_{ij}$ by performing integration over $q^2$ in the kinematically allowed
region. The averaging of $P_{ij}$ over $q^2$ is defined as
\bea
\label{e6628} 
\lla P_{ij} \rra = \ds \frac{\int_{4 m_\ell^2}^{(m_{\Lambda_b}-m_\Lambda)^2}
P_{ij} \ds \frac{d {\cal B}}{dq^2} dq^2}
{\int_{4 m_\ell^2}^{(m_{\Lambda_b}-m_\Lambda)^2} 
\ds \frac{d {\cal B}}{dq^2} dq^2}
\eea

In Figs. (1)--(5) we present the correlation of the averaged double--lepton
polarization asymmetries on the branching ratio for the $\Lambda_b \rar 
\Lambda \mu^+ \mu^-$ decay. From these figures we obtain the following
results.

\begin{itemize}
\item There exist regions of the new Wilson coefficients where $\lla
P_{LL}\rra$ departs considerably from the SM result when
${\cal B} (\Lambda_b \rar\Lambda \mu^+ \mu^-)$ is close to the SM
prediction.

\item $\lla P_{LT}\rra$ and $\lla P_{TL}\rra$ are both very sensitive to the
existence of the tensor and scalar interactions. We observe that $\lla
P_{LT}\rra$ exceeds the SM prediction
3 times and  $\lla P_{TL}\rra$  2 times, respectively. 
More essential than that, $\lla P_{LT}\rra$ as
well as $\lla P_{TL}\rra$ both change their signs when Wilson coefficients
vary in the allowed region. Such behaviors can serve as a good test for
establishing new physics beyond the SM.   

\item $\lla P_{NN}\rra$ and $\lla P_{TT}\rra$ are quite sensitive to the
existence of the vector interactions, especially to $C_{LR}$. In the
presence of this coefficient, the values of $\lla P_{NN}\rra$ and 
$\lla P_{TT}\rra$ can both exceed the SM result 3--4 times, and both changes
their sign when $C_{LR}$ varies in the allowed region. Therefore,
determination of the sign and magnitude of $\lla P_{NN}\rra$ and $\lla
P_{TT}\rra$ can give unambiguous information about the existence of the
new vector type interaction.
\end{itemize}

We do not present the correlation of and $\lla P_{LN}\rra$, 
$\lla P_{NL}\rra$, $\lla P_{NT}\rra$ and $\lla P_{TN}\rra$ on the branching
ratio, since their values are quite small. 

In Figs. (6)--(12) the correlation of $\lla P_{ij}\rra$ on branching ratio 
for the $\Lambda_b \rar\Lambda \tau^+ \tau^-$ decay are presented. Similar
to the $\Lambda_b \rar\Lambda \mu^+ \mu^-$ decay, we observe that several of
the double--lepton polarization asymmetries are very sensitive to the
existence of new physics and the presence of the new Wilson coefficients can
produce results that depart considerably from the SM prediction. More
precisely, we can briefly comment on the results as follows:

\begin{itemize}

\item All $\lla P_{ij}\rra$ are very sensitive to the existence of tensor
interactions with coefficients $C_T$ and $C_{TE}$.

\item The situation for the $\Lambda_b \rar\Lambda \tau^+ \tau^-$ decay is 
slightly different from the $\Lambda_b \rar\Lambda \mu^+ \mu^-$ decay, 
for which a considerable dependence on the existence of
the scalar type interaction is observed, especially for the coefficients
$C_{LRLR}$ and $C_{LRRL}$.

\item $\lla P_{LL}\rra$, $\lla P_{LT}\rra$, $\lla P_{TL}\rra$ and 
$\lla P_{TT}\rra$ exhibit strong dependence on vector interaction with
coefficients $C_{LR}$, while $\lla P_{LN}\rra$ and $\lla P_{NL}\rra$
do so on the vector interaction with coefficient $C_{LL}$.

\item $\lla P_{LL}\rra$ and  $\lla P_{NN}\rra$ change sign when the new
Wilson coefficients vary in the allowed region. For this reason, measurement
of magnitude and sign of the averaged double--lepton polarization asymmetry
can be a very useful tool in establishing new physics beyond the SM.
\end{itemize}
  
The following question we would like to discuss is the possibility of the
existence of such a region of the new Wilson coefficients for which 
the branching ratio coincides with SM result, while the double--lepton
polarizations do not, similar to the $\Lambda_b \rar \Lambda \mu^+ \mu^-$
decay.

From the relevant figures we see that, the  $\Lambda_b \rar\Lambda \tau^+
\tau^-$ decay is far more informative on this issue. There exist regions of
the new Wilson coefficients $C_{LR}$, $C_T$ and $C_{TE}$, where
double--lepton polarization asymmetries differ from the SM result, but
branching ratio coincides with that of the SM.

At the end of this section, let us discuss the problem of measurement of the
lepton polarization asymmetries in experiments. Experimentally, to measure an
asymmetry $\la P_{ij} \ra$ of the decay with the branching ratio ${\cal B}$
at $n \sigma$ level, the required number of events 
(i.e., the number of $B \bar{B}$ pair) are given by the expression
\bea
N = \frac{n^2}{{\cal B} s_1 s_2 \la P_{ij} \ra^2}~,\nnb
\eea
where $s_1$ and $s_2$ are the efficiencies of the leptons. Typical values of
the efficiencies of the $\tau$--leptons range from $50\%$ to $90\%$ for their
various decay modes (see for example \cite{R6625} and references therein), 
and the error in $\tau$--lepton 
polarization is estimated to be about $(10 \div 15)\%$ \cite{R6626}.
As a result, the error in measurement of the $\tau$--lepton asymmetries is of the
order of $(20 \div 30)\%$, and the error in obtaining the number of events is
about $50\%$.

From the expression for $N$ we see that, in order to observe the lepton
polarization asymmetries in $\Lambda_b \rar\Lambda \mu^+ \mu^-$ and
$\Lambda_b \rar\Lambda \tau^+ \tau^-$
decays at $3\sigma$ level, the minimum number of required events
are (for the efficiency of $\tau$--lepton we take $0.5$):

\begin{itemize}
\item for the $\Lambda_b \rar\Lambda \mu^+ \mu^-$ decay
\bea
N = \left\{ \begin{array}{ll}
2.0 \times 10^{6}  & (\mbox{\rm for} \lla P_{LL} \rra)~,\\
2.0 \times 10^{8}  & (\mbox{\rm for} \lla P_{LT} \rra=\lla P_{TL} \rra,
\lla P_{NN} \rra,\lla P_{TT} \rra)~,\end{array} \right. \nnb
\eea

\item for $\Lambda_b \rar\Lambda \tau^+ \tau^-$ decay
\bea
N = \left\{ \begin{array}{ll}
(4.0 \pm 2) \times 10^{9}  & (\mbox{\rm for} \lla P_{LT}\rra,
\lla P_{NN} \rra)~,\\   
(1.0 \pm 0.5) \times 10^{9}  & (\mbox{\rm for} \lla P_{TT} \rra)~,\\     
(2.0 \pm 1.0) \times 10^{11} & (\mbox{\rm for} \lla P_{LN} \rra,
\lla P_{NL} \rra)~,\\
(9.0 \pm 4.5) \times 10^{8} & (\mbox{\rm for} \lla P_{TL} \rra)~.
\end{array} \right.
\nnb
\eea
\end{itemize}

The number of $B \bar{B}$ pairs, that are produced
at B--factories and LHC \, are about $\sim 5\times 10^8$ and $10^{12}$,
respectively. As a result of a comparison of these numbers and $N$, we
conclude that, only $\lla P_{LL} \rra$ in the $\Lambda_b \rar\Lambda 
\mu^+ \mu^-$ decay and  $\lla P_{LT} \rra$, $\lla P_{NN} \rra$ and 
$\lla P_{TL} \rra$ in the $\Lambda_b \rar\Lambda \tau^+ \tau^-$ decay,
can be detectable at LHC. 

In summary, we present the most general analysis of the
double--lepton polarization asymmetries in the 
$\Lambda_b \rar\Lambda \ell^+ \ell^-$
decay using the most general, model independent form of the effective
Hamiltonian. The correlation of the averaged double--lepton
polarization asymmetries on the branching ratio have been studied. Our
results show that the study of double--lepton
polarization asymmetries can serve as good test for establishing
new physics beyond the SM. Moreover, we observe that there
exist regions of the new Wilson coefficients for which double--lepton
polarization asymmetries depart considerably from the SM, while the
branching ratio coincides with that of the SM predictions.

\newpage

\section*{Figure captions}
{\bf Fig. (1)} Parametric plot of the correlation between the averaged 
double--lepton polarization asymmetry $\la P_{LL} \ra$ 
and the branching ratio for the $\Lambda_b \rar \Lambda \mu^+ \mu^-$ decay, when 
both leptons are longitudinally polarized.\\ \\
{\bf Fig. (2)} The same as in Fig. (1), but for the averaged
double--lepton polarization asymmetry $\la P_{LT} \ra$.\\ \\
{\bf Fig. (3)} The same as in Fig. (1), but for the averaged
double--lepton polarization asymmetry $\la P_{TL} \ra$.\\ \\
{\bf Fig. (4)} The same as in Fig. (1), but for the averaged
double--lepton polarization asymmetry $\la P_{NN} \ra$.\\ \\
{\bf Fig. (5)} The same as in Fig. (1), but for the averaged
double--lepton polarization asymmetry $\la P_{TT} \ra$.\\ \\
{\bf Fig. (6)} The same as in Fig. (1), but for the 
$\Lambda_b \rar \Lambda \tau^+ \tau^-$ decay.\\ \\
{\bf Fig. (7)} The same as in Fig. (6), but for the averaged
double--lepton polarization asymmetry $\la P_{LN} \ra$.\\ \\
{\bf Fig. (8)} The same as in Fig. (6), but for the averaged
double--lepton polarization asymmetry $\la P_{NL} \ra$.\\ \\
{\bf Fig. (9)} The same as in Fig. (2), but for the 
$\Lambda_b \rar \Lambda \tau^+ \tau^-$ decay.\\ \\
{\bf Fig. (10)} The same as in Fig. (3), but for the
$\Lambda_b \rar \Lambda \tau^+ \tau^-$ decay.\\ \\
{\bf Fig. (11)} The same as in Fig. (4), but for the
$\Lambda_b \rar \Lambda \tau^+ \tau^-$ decay.\\ \\
{\bf Fig. (12)} The same as in Fig. (10), but for the averaged
double--lepton polarization asymmetry $\la P_{TT} \ra$.

\newpage

\begin{figure}
\vskip 1.5 cm
    \includegraphics{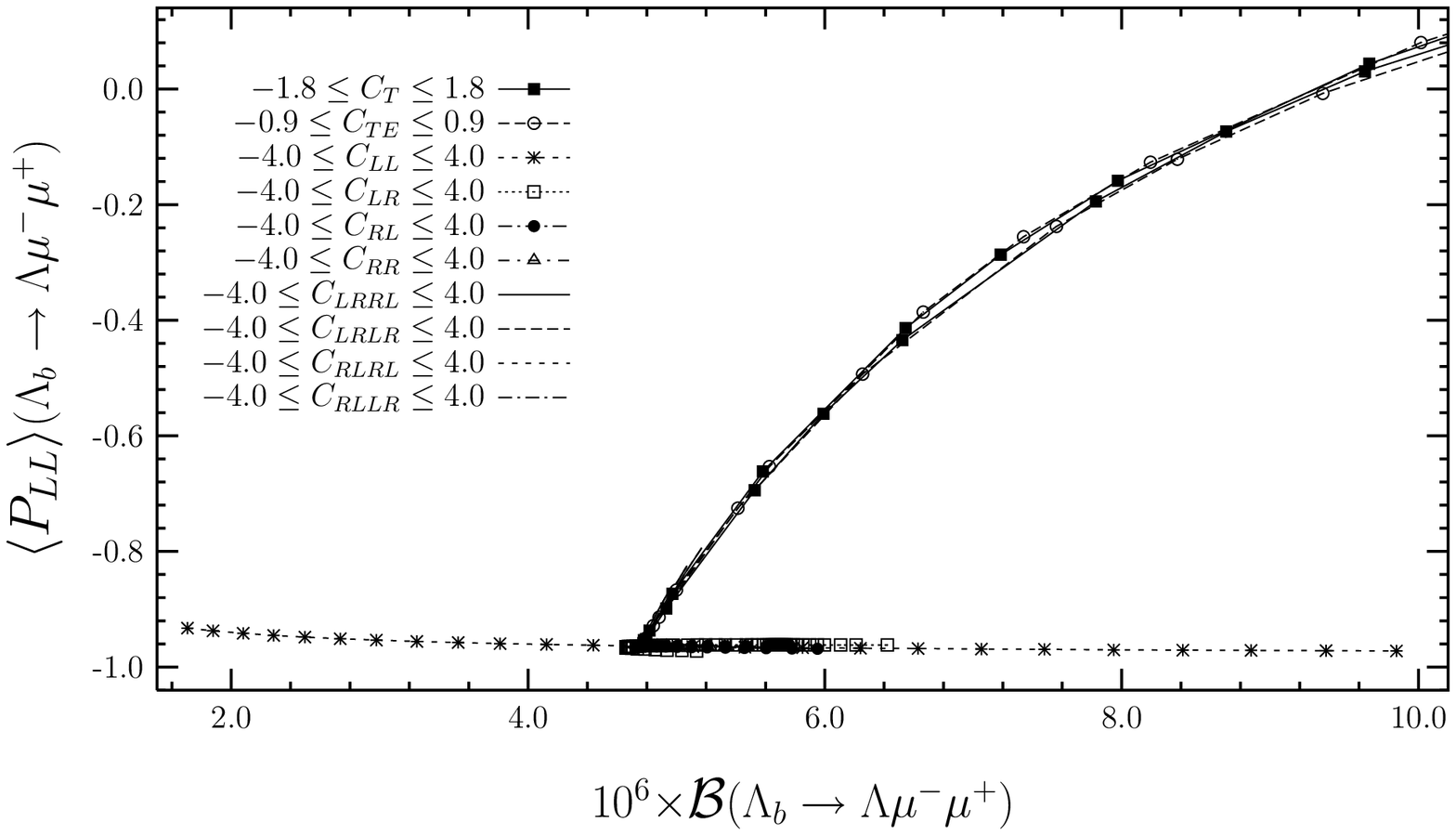}
\vskip 7.8cm
\caption{}
%\begin{center}
%{\bf Fig. 1--a}
%\end{center}
\end{figure}

\begin{figure}
\vskip 2.5 cm
    \includegraphics{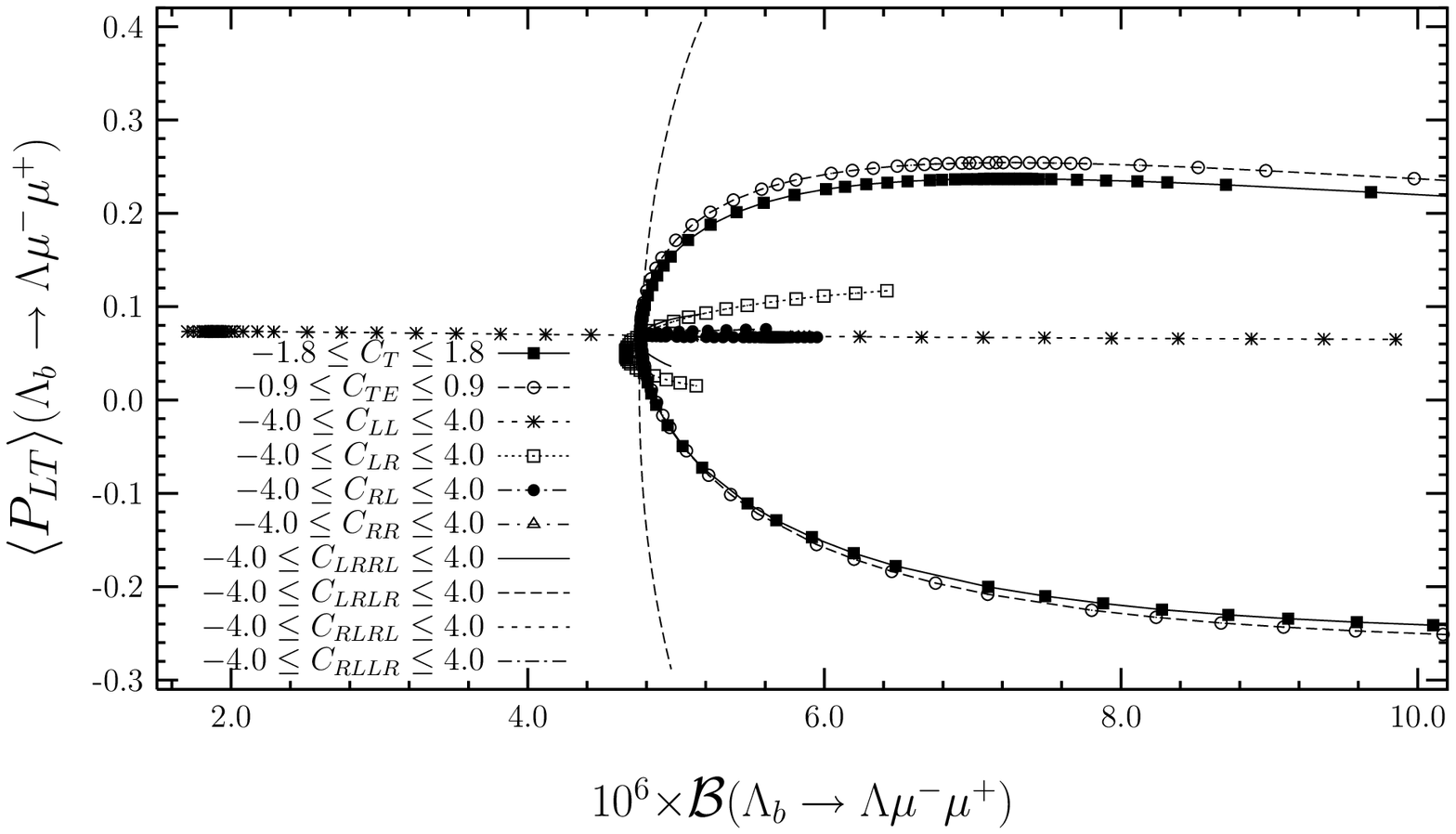}
\vskip 7.8 cm
\caption{}
%\begin{center}
%{\bf Fig. 1--b}
%\end{center}
\end{figure}

\begin{figure}
\vskip 1.5 cm
    \includegraphics{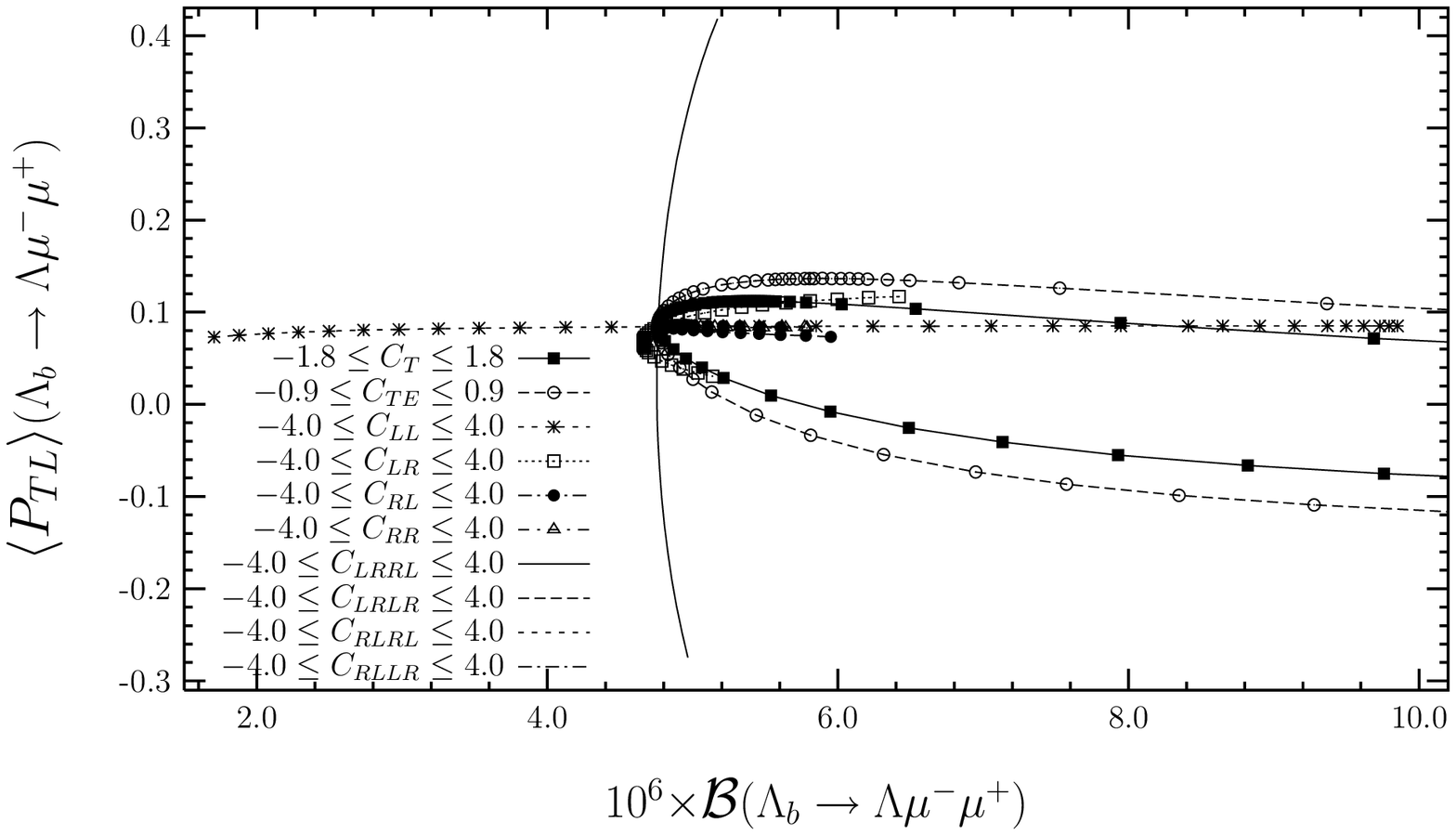}
\vskip 7.8cm
\caption{}
%\begin{center}
%{\bf Fig. 1--a}
%\end{center}
\end{figure}

\begin{figure}
\vskip 2.5 cm
    \includegraphics{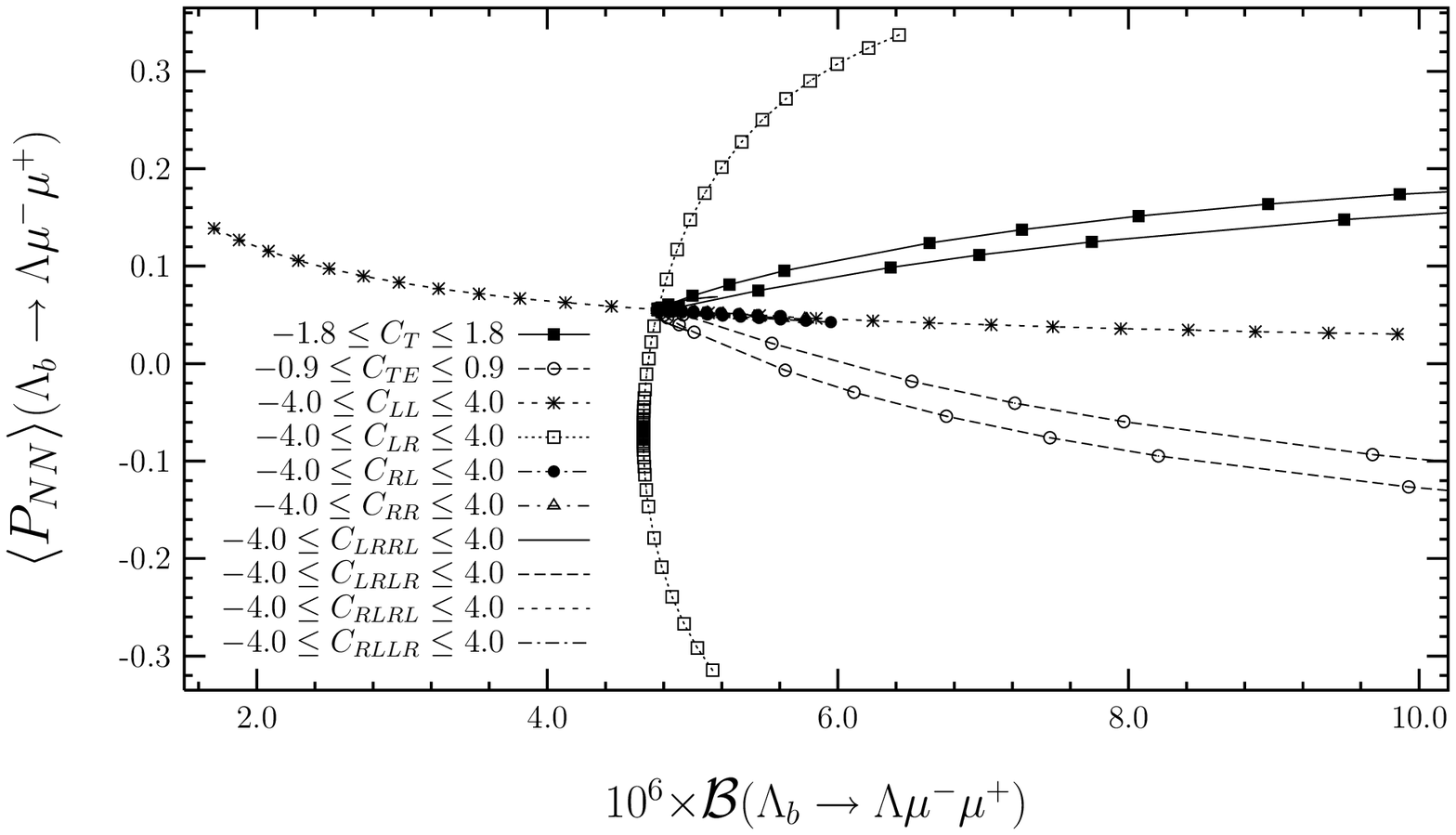}
\vskip 7.8 cm
\caption{}
%\begin{center}
%{\bf Fig. 1--b}
%\end{center}
\end{figure}

\begin{figure}
\vskip 2.5 cm
    \includegraphics{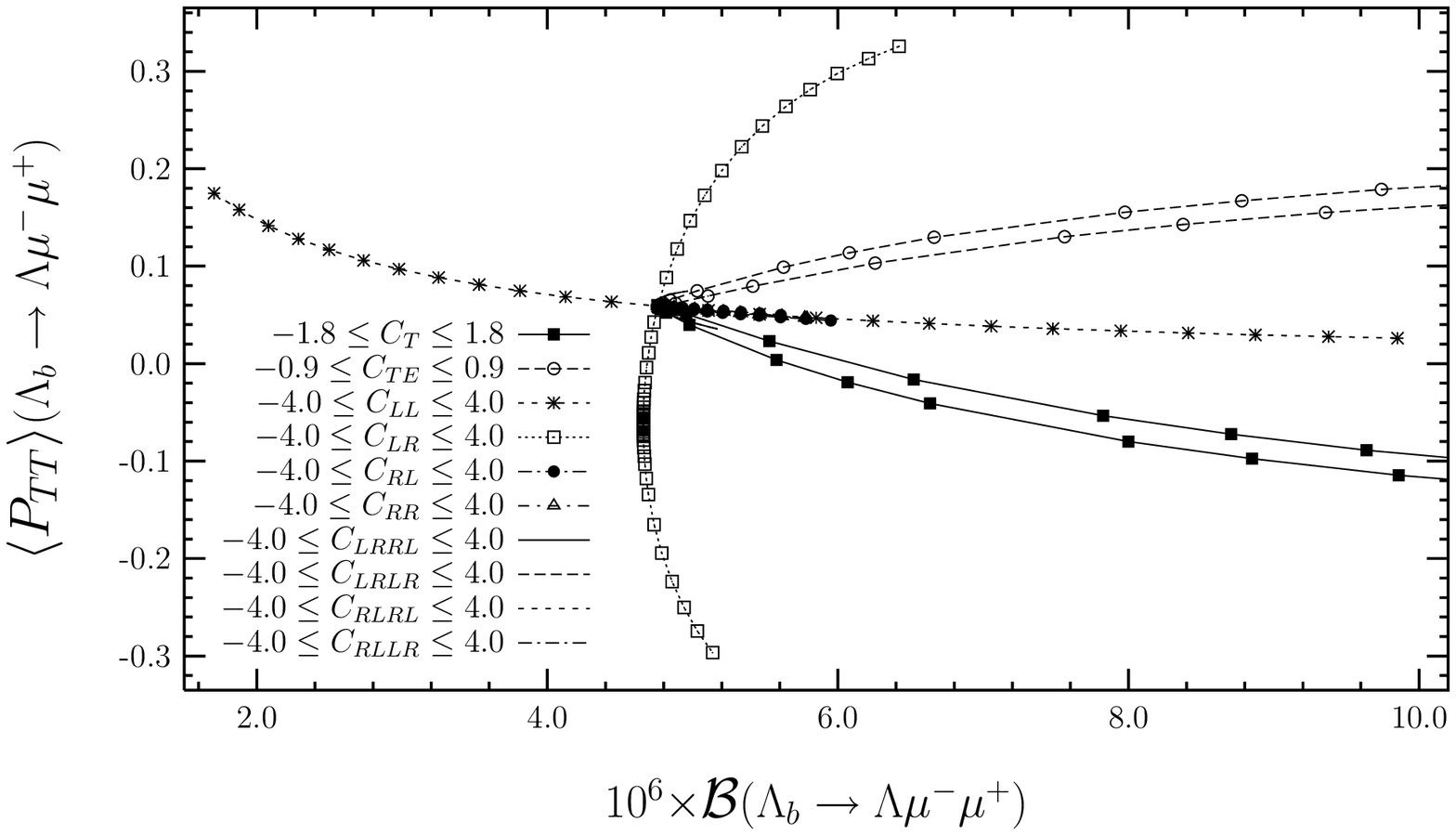}
\vskip 7.8 cm
\caption{}
%\begin{center}
%{\bf Fig. 1--b}
%\end{center}
\end{figure}

\begin{figure}
\vskip 1.5 cm
    \includegraphics{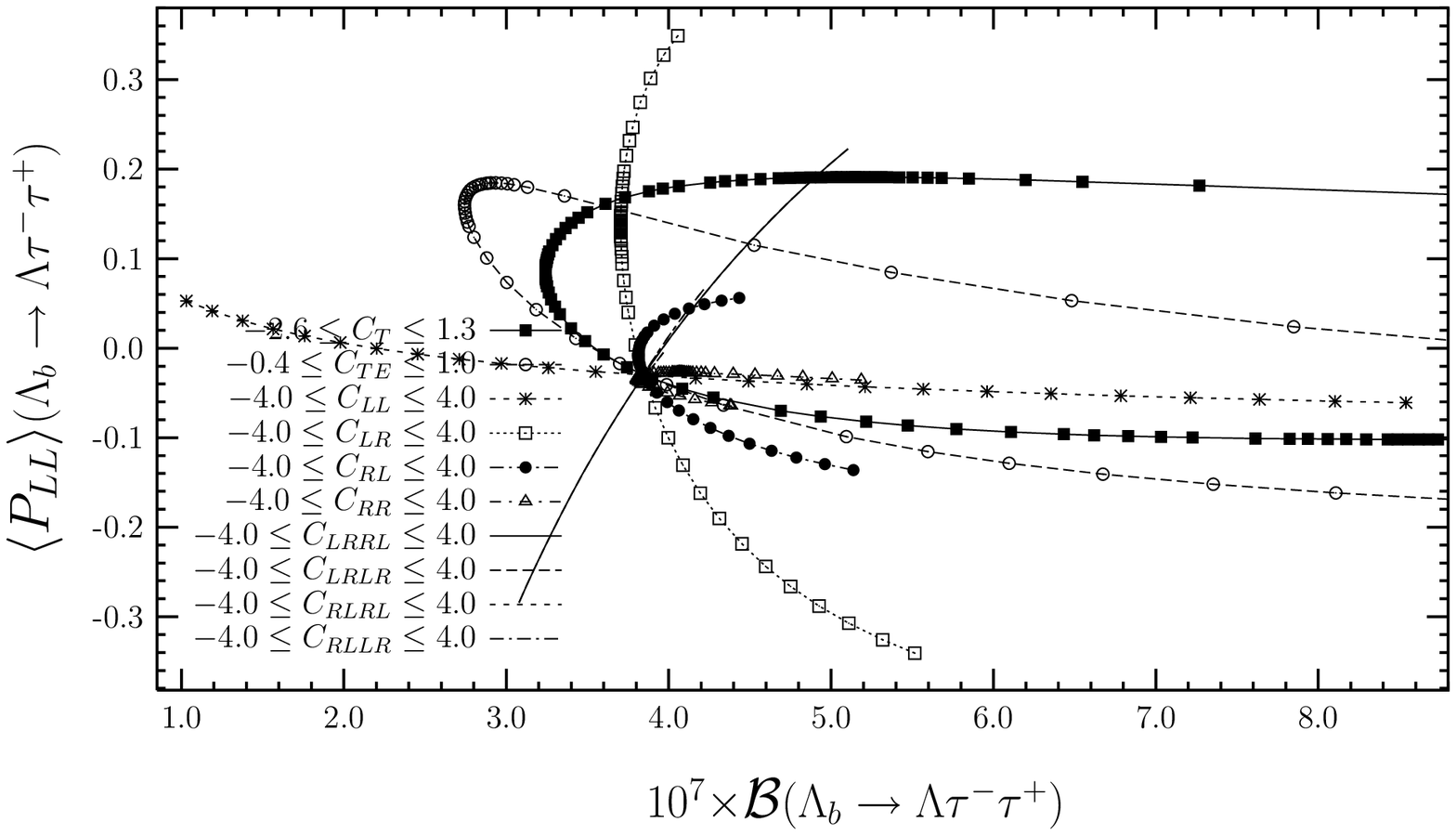}
\vskip 7.8cm
\caption{}
%\begin{center}
%{\bf Fig. 1--a}
%\end{center}
\end{figure}

\begin{figure}
\vskip 2.5 cm
    \includegraphics{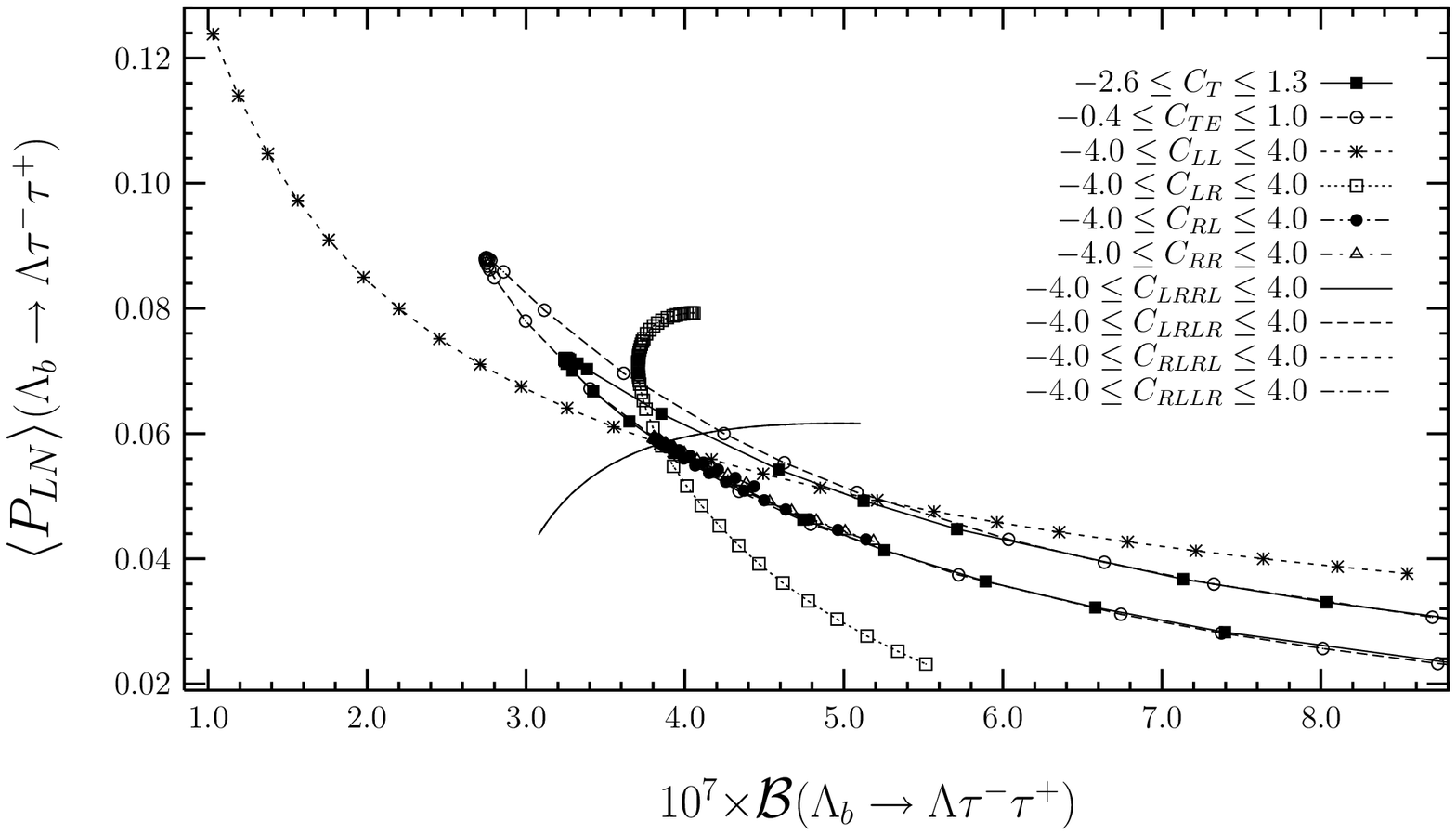}
\vskip 7.8 cm
\caption{}
%\begin{center}
%{\bf Fig. 1--b}
%\end{center}
\end{figure}

\begin{figure}
\vskip 1.5 cm
    \includegraphics{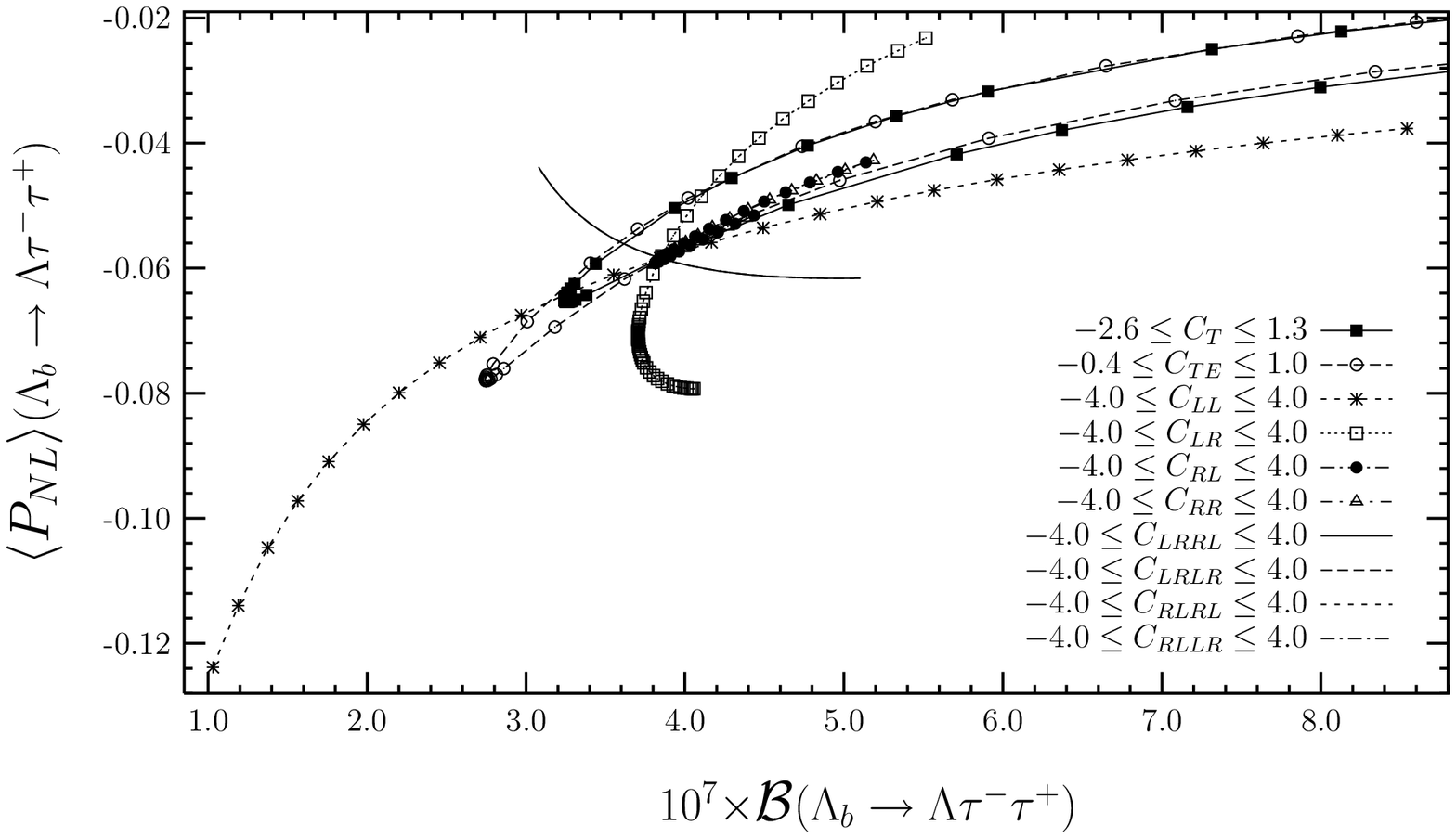}
\vskip 7.8cm
\caption{}
%\begin{center}
%{\bf Fig. 1--a}
%\end{center}
\end{figure}

\begin{figure}
\vskip 2.5 cm
    \includegraphics{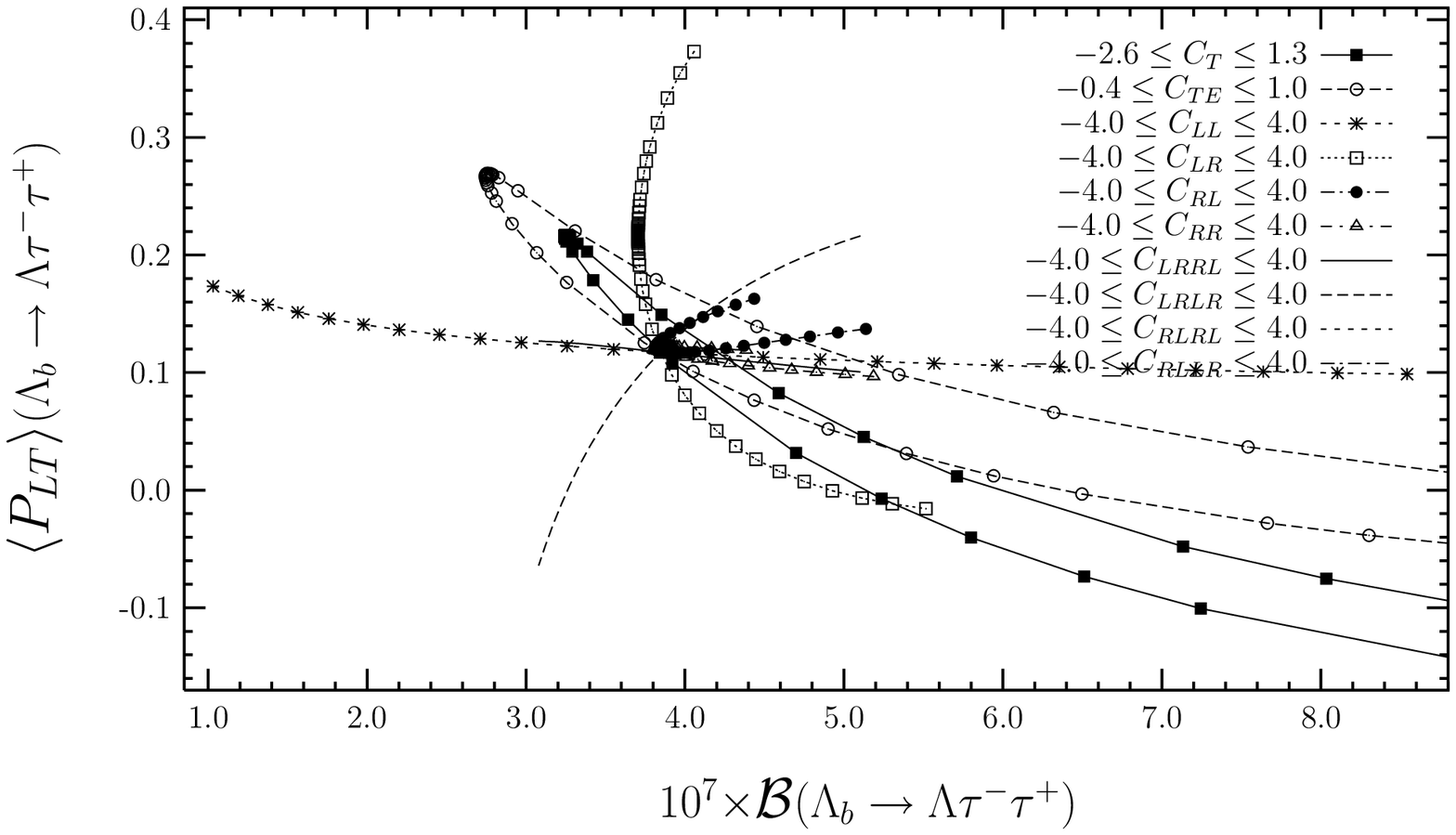}
\vskip 7.8 cm
\caption{}
%\begin{center}
%{\bf Fig. 1--b}
%\end{center}
\end{figure}

\begin{figure}
\vskip 1.5 cm
    \includegraphics{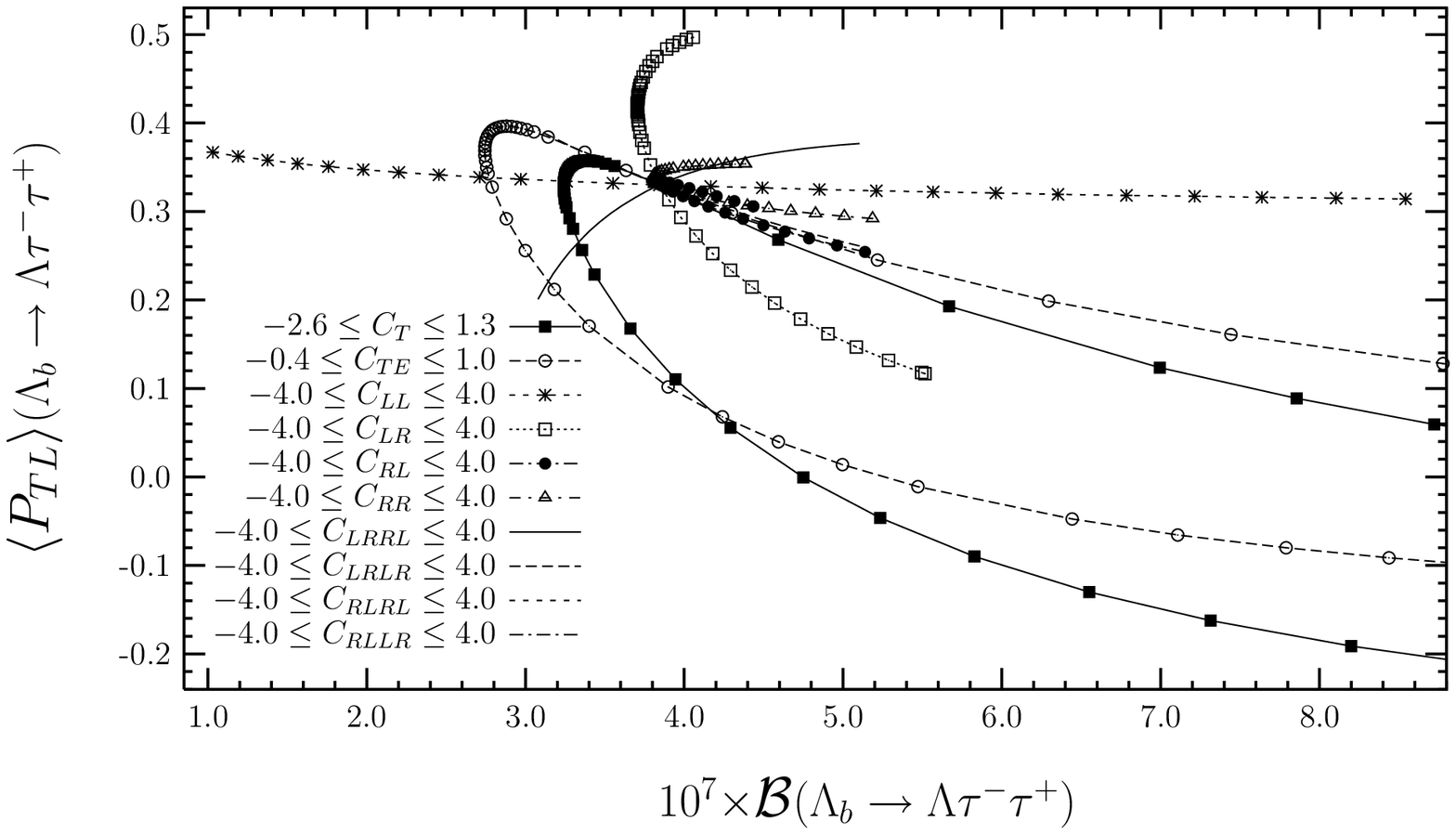}
\vskip 7.8cm
\caption{}
%\begin{center}
%{\bf Fig. 1--a}
%\end{center}
\end{figure}

\begin{figure}
\vskip 2.5 cm
    \includegraphics{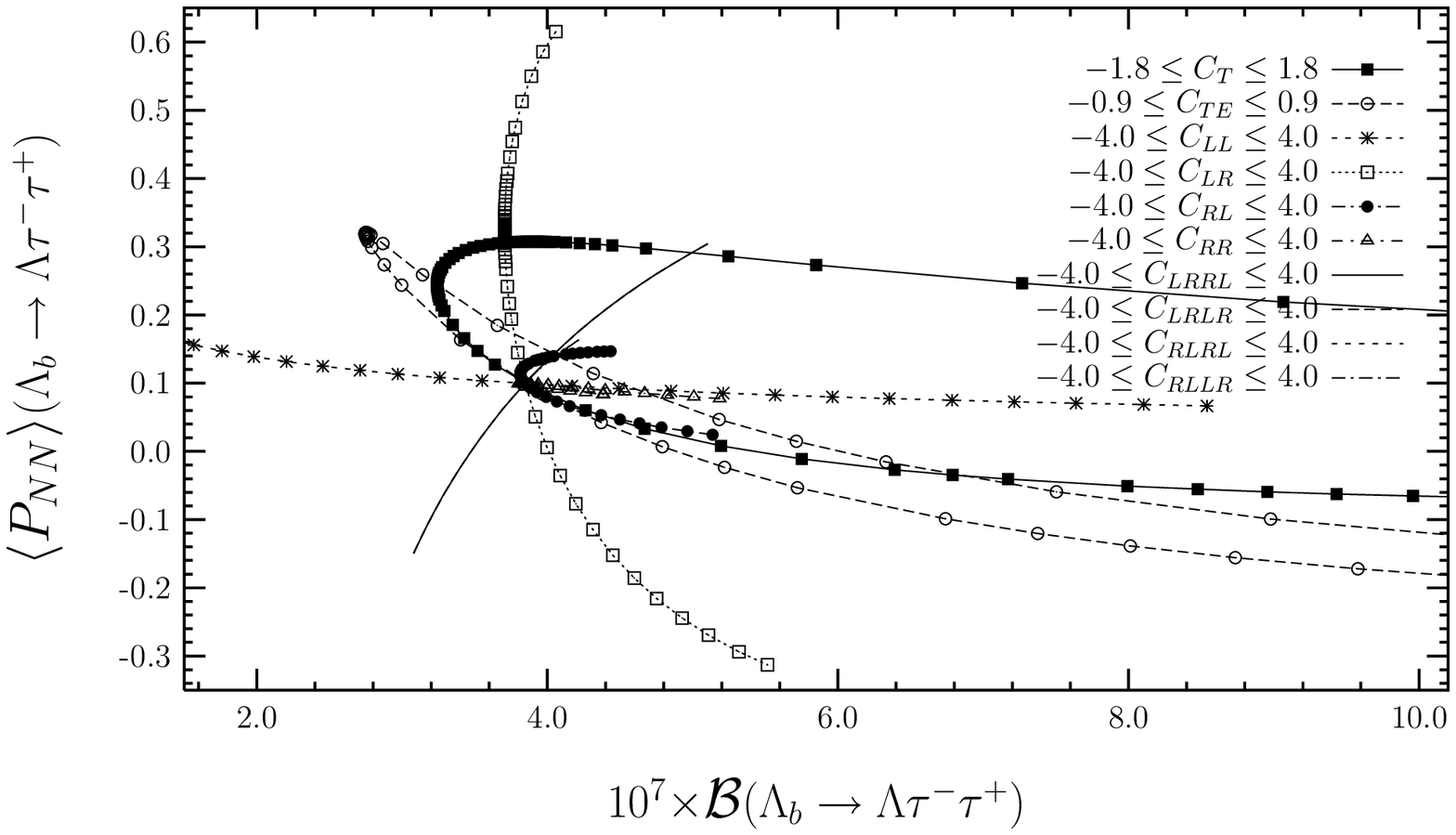}
\vskip 7.8 cm
\caption{}
%\begin{center}
%{\bf Fig. 1--b}
%\end{center}
\end{figure}

\begin{figure}
\vskip 1.5 cm
    \includegraphics{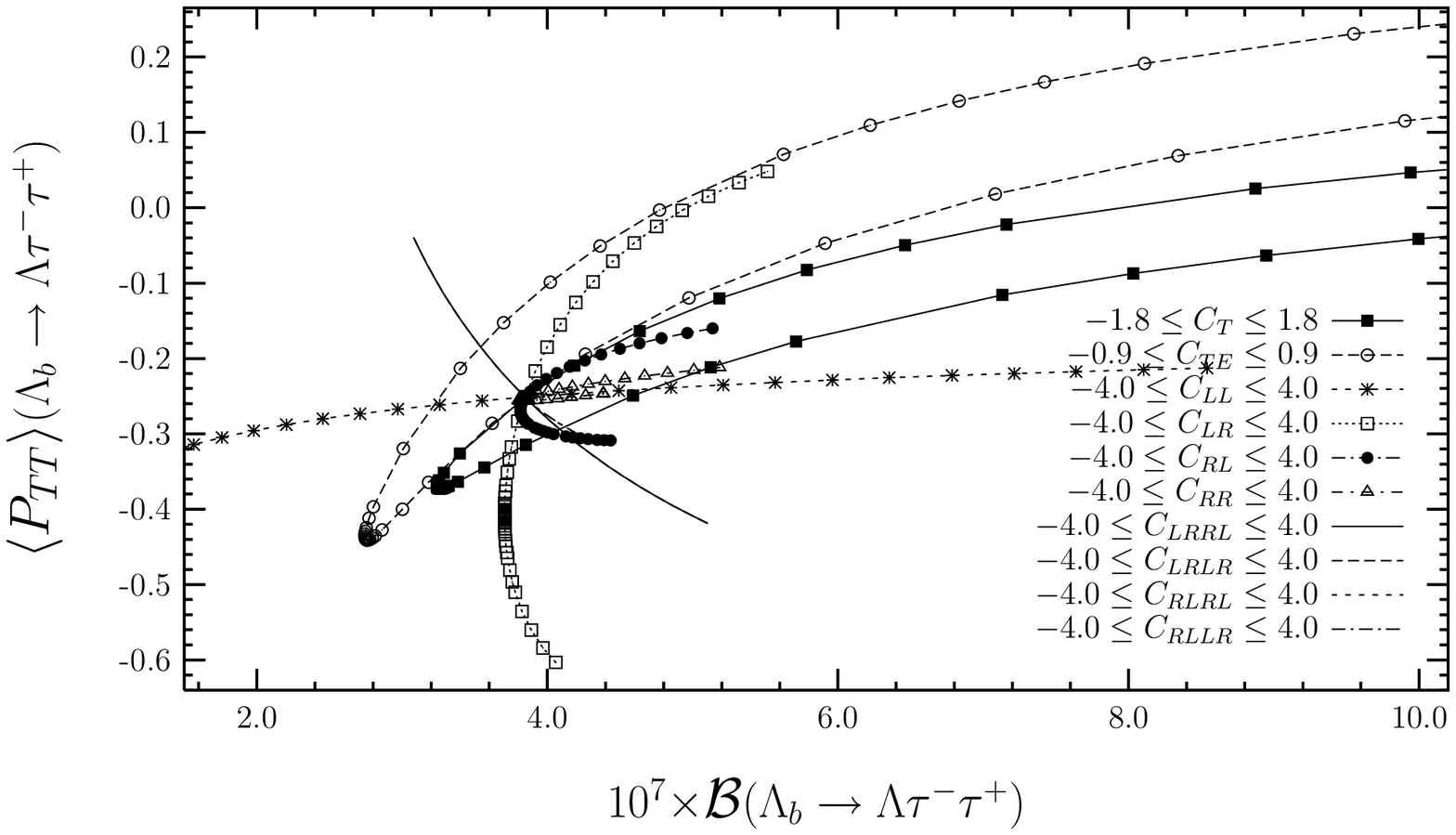}
\vskip 7.8cm
\caption{}
%\begin{center}
%{\bf Fig. 1--a}
%\end{center}
\end{figure}

\end{document}